\documentstyle[prd,aps,twocolumn]{revtex}
\begin{document}
\draft

\twocolumn[\hsize\textwidth\columnwidth\hsize\csname @twocolumnfalse\endcsname
\title{Quantum Weakdynamics as an $SU(3)_I$ Gauge Theory: Grand Unification of Strong and Electroweak Interactions}
\author{Heui-Seol Roh\thanks{e-mail: hroh@nature.skku.ac.kr}}
\address{BK21 Physics Research Division, Department of Physics, Sung Kyun Kwan University, Suwon 440-746, Republic of Korea}
\date{\today}
\maketitle
\begin{abstract}
Quantum weakdynamics (QWD) as an $SU(3)_I$ gauge theory with the $\Theta$ vacuum term
is considered to be the unification of the electroweak interaction as an $SU(2)_L \times
U(1)_Y$ gauge theory. The grand unification of $SU(3)_I \times SU(3)_C$ beyond the
standard model $SU(3)_C \times SU(2)_L \times U(1)_Y$ is established by the group
$SU(3)_I$. The grand unified interactions break down to weak
and strong interactions at a new grand unification scale $10^{3}$ GeV,
through dynamical spontaneous symmetry breaking (DSSB); the weak
and strong coupling constants are the same, $\alpha_i = \alpha_s \simeq 0.12$, at this scale. DSSB is
realized by the condensation of scalar fields, postulated to be spatially longitudinal
components of gauge bosons, instead of Higgs particles.
Quark and lepton family generation, the Weinberg angle $\sin^2 \theta_W =
1/4$, and the Cabbibo angle $\sin \theta_C = 1/4$ are predicted. The electroweak coupling
constants are $\alpha_z = \alpha_i/3$, $\alpha_w = \alpha_i/4$, $\alpha_y =
\alpha_i/12$, and $\alpha_e = \alpha_i/16 \simeq 1/137$; There are symmetric isospin interactions.
\end{abstract}
\maketitle
] \narrowtext

\section{Introduction}
\label{sec1}

The standard model (SM) $SU(3)_C \times SU(2)_L \times U(1)_Y$ gauge theory
\cite{Glas,Frit} is quite successful for the phenomenology of elementary particle
physics. However, there are still many unexplained phenomena in the SM: for instances,
many free parameters, three family generations for leptons and quarks, matter mass
generation, the Higgs problem or vacuum problem, dynamical spontaneous symmetry breaking
(DSSB) beyond spontaneous symmetry breaking, neutrino mass problem, etc. In order to
resolve these problems, grand unified theories (GUTs) were proposed \cite{Geor}.
Nevertheless, grand unification of the strong and electroweak interactions is not
complete, and GUTs also have model dependent problems: the hierarchy problem, proton decay,
and the Weinberg angle are problems in $SU(5)$ gauge theory \cite{Geor}. On the other
hand, two important observations can be made. One is that Higgs particles have not been
observed yet; this suggests DSSB rather than spontaneous symmetry breaking. The other one
is that the experimental strong coupling constant $\alpha_s \simeq 0.12$ at the energy
scale of the intermediate $Z^0$ vector boson mass \cite{Hinc} and the experimental
weak coupling constant $\alpha_w \simeq 0.03$ at the energy scale of the $W^{\pm}$ intermediate
vector boson mass \cite{Bard}; the two coupling constants have the same value
$\alpha_i = \alpha_s \simeq 0.12$ around the $10^3$ GeV energy scale if an $SU(3)_I$ gauge
theory with the coupling constant $\alpha_i$ for the weak force is adopted. These
phenomena strongly suggest that all the theoretical problems and experimental facts
may thus easily be resolved if quantum weakdynamics (QWD) as an $SU(3)_I$ gauge theory
for the weak force is broken down to the Glashow-Weinberg-Salam (GWS) model, the
$SU(2)_L \times U(1)_Y$ theory \cite{Glas} through DSSB. The aim of this paper
accordingly is to propose that QWD as an $SU(3)_I$ gauge theory for the weak force provides
hints for the challenging problems and ways to unify the weak and strong force
systematically: QWD is analogous to quantum chromodynamics (QCD) as an $SU(3)_C$
gauge theory for the strong force. Many free parameters in the SM must be understood
in the context of the grand unification scheme. More precisely, QWD as an $SU(3)_I$
gauge theory is proposed to be the unification of the $SU(2)_L \times U(1)_Y$ electroweak
theory \cite{Glas}, and the grand unification of QCD and QWD is proposed as the
unification of weak and strong interactions \cite{Frit,Roh3,Roh31} beyond the SM. The
proposed group chain is thus given by $H \supset SU(3)_I \times SU(3)_C$ for grand
unification, weak, and strong interactions respectively. The grand unified group $H$
of the group $SU(3)_I \times SU(3)_C$ beyond the SM, $SU(3)_C \times SU(2)_L \times
U(1)_Y$, provides coupling constants $\alpha_w = \alpha_s \simeq 0.12$ at a new grand
unification scale around $10^{3}$ GeV, which might be the resolution of the hierarchy
problem of the conventional grand unification scale $10^{15}$ GeV \cite{Geor}. QWD
provides plausible explanations for the Weinberg angle, the Cabbibo angle, quark and
lepton families, modification, etc.
This scheme can be substantiated through further quantum tests; it gives rise to
several predictions such as the relation with inflation, the analogy between weak
and strong force, the breaking of discrete symmetries, etc.  The present work is
restricted to the real four dimensions of spacetime without considering supersymmetry
or higher dimensions. This work is based on phenomenology below the grand unification
scale: electroweak and strong interactions.

As a step toward the grand unification of fundamental forces or
toward the systematic description of the evolution of the universe, a
new grand unification scale of the $SU(3)_I \times SU(3)_C$
symmetry around $10^{3}$ GeV is necessary. QWD as an $SU(3)_I$
gauge theory provides the unification of electroweak interaction being
an $SU(2)_L \times U(1)_Y$ gauge theory \cite{Glas}. This
scheme resolves the free parameters in the GWS model. In order to
show that electroweak interactions stem from an $SU(3)_I$ gauge
theory, the roles of scalar fields, parameterized by spatially
longitudinal components of gauge bosons, are emphasized instead of
roles of Higgs particles: unsatisfactory factors of the Higgs
mechanism \cite{Higg} can be overcome in this scheme. The
essential point is that the system at high energies experiences
the stage of DSSB and the effective coupling constant acquires
the dimension of inverse energy square due to massive gauge
bosons through DSSB; the effective coupling constant chain due to
massive gauge bosons is $G_H \supset G_F \times G_R$ for grand
unification, weak, and strong interactions respectively. The DSSB
mechanism is adopted to all the interactions characterized by
gauge invariance, the physical vacuum problem, and discrete symmetry
breaking; the DSSB mechanism is applied to strong interactions having
analogous features \cite{Roh3,Roh31}. The DSSB of local
gauge symmetry and global chiral symmetry triggers the (V+A)
current anomaly. This study suggests that sextet isospin states
in two octets of triplet isospin combinations for QWD can produce
the three family generations of leptons and quarks. DSSB is closely
related to the breaking of discrete symmetries, parity (P),
charge conjugation (C), charge conjugation and parity (CP) and
time reversal (T). Photons are regarded as massless gauge bosons,
Nambu-Goldstone (NG) bosons \cite{Namb}, appearing during DSSB.
The quark or lepton mass is generated as the DSSB of gauge
symmetry and discrete symmetries, which is motivated by the
parameter $\Theta$ representing the surface term. Fermion mass
generation introduces the common features of constituent particle
mass, the dual Meissner effect, and hyperfine structure. The $\Theta$
term plays important roles in the DSSB mechanism of the gauge
group and in the quantization of the fermion space and vacuum
space.

This paper is organized as follows. Section \ref{sec2} proposes
QWD as an $SU(3)_I$ gauge symmetry and describes the common
features of DSSB. Section \ref{sec3} explicitly shows the
generation of the $SU(2)_L \times U(1)_Y$ gauge symmetry from the
$SU(3)_I$ gauge symmetry. In Sect. \ref{sec4}, the grand
unification of $SU(3)_I \times SU(3)_C$ introduces the
unified coupling constant at a new grand unification scale.
The fermion mass generation mechanism is addressed as being the result of
the breaking of gauge and chiral symmetries in Sect. \ref{sec5}.
The $\Theta$ constant and quantum numbers are discussed in Sect.
\ref{sec6}. Section \ref{sec7} describes a comparison of QWD,
GWS, and GUT. Section \ref{sec8} is devoted to conclusions.

\section{Quantum Weakdynamics as an $SU(3)_I$ Gauge Theory}
\label{sec2}

The generation of the electroweak $SU(2)_L \times U(1)_Y$ theory from QWD as the
$SU(3)_I$ gauge theory has several implications. It suggests the DSSB mechanism
initiated by the (V+A) current anomaly and predicts several free parameters such as
the Weinberg angle, the Cabbibo angle, isospin and electric charge quantization,
modification to the Higgs mechanism, the three family generations of leptons and quarks,
fermion mass generation, etc.

DSSB is one of the underlying principles whose principal application is the
electroweak theory; remarkably, this unifies weak interactions with electromagnetic
interactions in a single larger gauge theory. Here, the DSSB of the weak force
from an $SU(3)_I$ gauge theory to an $SU(2)_L \times U(1)_Y$ electroweak theory, the
GWS model \cite{Glas}, is addressed as well as the modification of the Higgs mechanism
\cite{Higg}; the phase transition of electroweak interactions takes place through
the condensation of scalar fields, which are postulated to be spatially longitudinal
components of gauge bosons, instead of Higgs particles.
This scheme uses dynamical symmetry breaking without having to introduce elementary scalar
particles; this idea, which aims to have DSSB with gauge interactions alone, is
similar to the technicolor model \cite{Suss} in this sense. Discrete symmetry violation
occurs as the result of DSSB. Photons are regarded as massless gauge bosons
\cite{Namb} indicating DSSB. An analogy of superconductivity is expected as a consequence of
the condensation of scalar fields.

In this section, the common characteristics of DSSB are discussed
following the introduction of QWD: the introduction of QWD, the
$\Theta$ vacuum as weak CP violation, DSSB through the
condensation of scalar fields, the Fermi weak interaction constant and massive
gauge boson, and the breaking of discrete symmetries are
addressed.

\subsection{Weak Interactions as an $SU(3)_I$ Gauge Theory}

QWD is proposed as an $SU(3)_I$ gauge theory, which is an exact copy of QCD in the form
apart from the fermion mass term;
triplet isospin sources and octet gauge bosons are introduced.
A mass term violating gauge invariance is not included but is generated through DSSB.
Natural units with $\hbar = c = k = 1$ are used for convenience throughout this paper
unless otherwise specified.

The Dirac Lagrangian density apart from the mass term
${\cal L} = \bar \psi i \gamma^\mu \partial_\mu \psi$
leads to the local $SU(3)_I$ gauge theory invariant under the local transformation
$\psi \rightarrow e^{i \vec \lambda \cdot \vec a /2} \psi$
where $\vec \lambda$ are the $SU(3)_I$ generators and $\vec a$ are spacetime parameters.
An $SU(3)_I$ gauge theory for the weak interactions can then be established and
the $SU(3)_I$ gauge invariant Lagrangian density is the same, apart from the mass term,
as the Lagrangian density of QCD \cite{Frit} in the form
\begin{equation}
\label{qgsu}
{\cal L} = - \frac{1}{2} Tr F_{\mu \nu} F^{\mu \nu} + \bar \psi i \gamma^\mu D_\mu  \psi
\end{equation}
where $\psi$ stands for the spinor field and $D_\mu = \partial_\mu - i g_i G_\mu$ stands
for the covariant derivative with the coupling constant $g_i$.
The field strength tensor is given by
$F_{\mu \nu} = \partial_\mu G_\nu - \partial_\nu G_\mu - i g_i [G_\mu, G_\nu]$,
where spinors carry gauge fields denoted by $G_\mu = \sum_{a=0}^8 G_\mu^a \lambda^a /2, a = 0, ..,8$
with generators $\lambda^a$.
The Gell-Mann matrices satisfy the commutation relation
$[\lambda_k, \lambda_l ] = 2 i \sum_m c_{klm} \lambda_m$
where $c_{klm}$ are the structure constants of the $SU(3)_I$ group.
The fine structure constant $\alpha_i$ in QWD is defined analogously to the fine structure constant
$\alpha_e = e^2/ 4 \pi $ in QED:
$\alpha_i = g_i^2 / 4 \pi$
where $\alpha_i$ is dimensionless.

Three intrinsic isospin (isotope) charges $(A, B, C)$ form
the fundamental representation of the $SU(3)_I$ symmetry group.
Fermions are combinations of three particles with triplet isospins
which produce a decuplet, two octets, and a singlet gauge bosons;
$3 \otimes 3 \otimes 3 = 10 \oplus 8 \oplus 8 \oplus 1$.
Interactions between fermions may be described by two body
interactions. The set of unitary $3 \times 3$ matrices with
$\textup{det} \ U = 1$ forms the group $SU(3)_I$ whose fundamental
representation is a triplet. The eight gauge bosons in the octet
are for example constructed by a matrix $\sum^8_1 \lambda_k G^k$:
\begin{equation}
\label{suth}
\left ( \begin{array}{ccc} G_3 + G_8/\sqrt{3} & G_1 - iG_2 & G_4 -
i G_5 \\ G_1 + iG_2 & - G_3 +  G_8/\sqrt{3} & G_6 - i G_7 \\ G_4 +
i G_5 & G_6 + i G_7 & - 2  G_8/\sqrt{3}
\end{array} \right )
\end{equation}
where the two diagonal gauge bosons are $G_3 = (A \bar A - B \bar B)/\sqrt{2}$
and $G_8 = (A \bar A + B \bar B - 2 C \bar C)/\sqrt{6}$.

The six off-diagonal gauge bosons are accordingly represented by
\begin{eqnarray}
\label{glms} A \bar B & = & (G_1 - i G_2)/\sqrt{2}, \ A \bar C = (G_4 - i
G_5)/\sqrt{2},  \nonumber \\ B \bar C & = & (G_6 - i G_7)/\sqrt{2}, \ B \bar A  = (G_1
+ i G_2)/\sqrt{2},  \nonumber \\ C \bar A & = & (G_4 + i G_5)/\sqrt{2}, \ C \bar B =
(G_6 + i G_7)/\sqrt{2} .
\end{eqnarray}
The isospin singlet is symmetric:
\begin{equation}
G_0 = (A \bar A + B \bar B + C \bar C)/ \sqrt{3} .
\end{equation}
It will be later realized that $G_0$ is weak gauge boson with isospin zero,
$G_1 \sim G_3$ are weak gauge bosons with isospin one, and
$G_4 \sim G_8$ are weak gauge bosons with isospin two.

The conserved quantity $Q$ as a total electric charge is quantized in terms of Gell-Mann-Nishijima formula \cite{Gell}
as the subgroup $U(1)_e$ of the $SU(3)_I$ gauge symmetry via the $SU(2)_L \times U(1)_Y$ gauge group.
The corresponding charge operator $\hat Q$ is defined by
\begin{equation}
\label{chqu}
\hat Q = \hat I_{3} + \hat Y/2
\end{equation}
where $\hat I_{3}$ is the third component of the isospin operator $\hat I$ and $\hat Y = \hat B - \hat L$ is
the hypercharge operator $\hat Y$ with the baryon number operator $\hat B$ and lepton number operator $\hat L$.

\subsection{$\Theta$ Vacuum: Weak CP Asymmetry}

The (V+A) current anomaly is taken into account in analogy with
the axial current anomaly \cite{Adle}, which is linked to the
$\Theta$ vacuum in QCD as a gauge theory \cite{Hoof2,Roh3}. The
normal vacuum is unstable and the tunneling mechanism is possible
between all possible vacua. The true vacuum must be a
superposition of the various vacua, each belonging to some
different homotopy class. The effective $\bar \Theta$ term in QCD
involves both the bare $\Theta$ term relevant for pure QCD vacuum
and the phase of the quark mass matrix relevant for electroweak
effects. Only the latter is here considered as the $\Theta$
vacuum responsible for the fermion mass and is added as a single,
additional non-perturbative term to the QWD Lagrangian density
\begin{equation}
\label{thet}
{\cal L}_{QWD} = {\cal L}_{P} + \Theta \frac{g_i^2}{16 \pi^2} Tr F^{\mu \nu} \tilde F_{\mu \nu},
\end{equation}
where ${\cal L}_{P}$ is the perturbative Lagrangian density
(\ref{qgsu}), $F^{\mu \nu}$ is the field strength tensor, and
$\tilde F_{\mu \nu}$ is the dual of the field strength tensor.
Since the $F \tilde F$ term is a total derivative, it does not
affect the perturbative aspects of the theory. Such a term in the
QWD Lagrangian represents the (V+A) current anomaly, violates CP,
T, and P symmetries, and causes the DSSB of local gauge symmetry
and global chiral symmetry. Since the Lagrangian density for weak
interactions (\ref{qgsu}) is completely symmetric under the
$SU(3)_I$ gauge transformation, the conservation of isospin
degrees of freedom holds. Spatially longitudinal components of the gauge fields
parameterized by scalar fields condense and then subtract the
vacuum energy in broken phase as the system expands; scalar
fields replace the roles of Higgs bosons in the electroweak interactions.
Details are given in the following subsections.

\subsection{Dynamical Spontaneous Symmetry Breaking}

DSSB requires the $\Theta$ term and scalar fields. Scalar fields with spin zero,
which replace the roles of Higgs particles, are postulated as spatially longitudinal components of gauge fields.
Therefore, scalar fields always have the same symmetry as the gauge symmetry of gauge fields.
DSSB is triggered by the $\Theta$ term in equation (\ref{thet}) as the surface term, which demands
the boundary condition breaking the discrete symmetries.
Due to the $\Theta$ term, some of the scalar fields condense. There are two types of transverse gauge fields
which are left-handed (or vector) gauge fields and right-handed (or axial vector) gauge fields.
There are also two types of scalar fields which are real scalar fields and pseudo-scalar fields.
Among the scalar fields, pseudo-scalar fields condense while real scalar fields remain.
During this process, continuous and discrete symmetries are simultaneously broken.
This mechanism makes gauge bosons massive
and massive gauge bosons break the discrete symmetries in fermion spectra.
Masses of fermions and bosons are thus acquired as a consequence of DSSB.
In fact, gauge bosons become less massive and fermions become more massive as the condensation increases.
In the phase transition from the $SU(3)_I$ to the $SU(2)_L \times U(1)_Y$ gauge symmetry,
more massive gauge bosons with isospin $2$ become less important and less massive gauge bosons with
isospin $1$ become more important.
The scale of the gauge boson mass is related to the scale of the vacuum energy.
Left-handed and right-handed fermions are classified through the weak phase transition.
Overall, all these processes simultaneously and dynamically take place and both continuous and discrete symmetries are
spontaneously broken. In summary, DSSB mechanism involves the mass generation of fermions and bosons,
discrete symmetry breaking in fermions and bosons, and continuous symmetry breaking in scalar and gauge fields.

A fermion possesses three intrinsic isospin degrees of freedom;
in group theoretical language, $3 \otimes 3 \otimes 3 = 10 \oplus
8 \oplus 8 \oplus 1$. Two octets are mixed to form leptons and
quarks, which is discussed in the following section. In this
scheme, masses of the gauge bosons are reduced due to the
condensation of scalar fields, which are postulated as spatially
longitudinal components of gauge bosons. This is very analogous
to the mass generation of electroweak interactions through the
Higgs mechanism \cite{Higg}, but it is different in that the
gauge boson mass decreases as the condensation increases as shown
in the following subsection. Gauge fields are generally
decomposed by charge non-singlet-singlet symmetries on the one
hand and by even-odd discrete symmetries on the other hand: they
have dual properties in charge and discrete symmetries.  The
Higgs particles are not necessary in this case, since spatially
longitudinal components of the gauge bosons play the same role as
the Higgs particles. This scheme introduces dynamical symmetry
breaking without any free parameters except the weak coupling
constant. The $SU(3)_I$ gauge symmetry is spontaneously broken to
$SU(2)_L \times U(1)_Y$ gauge symmetry by the condensation of
scalar fields, postulated as spatially longitudinal components of
gauge fields. DSSB consists of two simultaneous mechanisms; the
first mechanism is the explicit symmetry breaking of the gauge
symmetry, which is represented by the isospin factor $i_f$ and
the weak coupling constant $g_i$, and the second mechanism is the
spontaneous symmetry breaking of the gauge fields, which is
represented by the condensation of pseudo-scalar fields.

Effective gauge boson interactions can, from the Lagrangian density (\ref{thet}), written by
\begin{equation}
\label{grac}
\triangle {\cal L}^e = - \frac{1}{2} Tr F_{\mu \nu} F^{\mu \nu} + \Theta
\frac{g_i^2}{16 \pi^2} Tr F^{\mu \nu} \tilde F_{\mu \nu} .
\end{equation}
Spatially longitudinal components of the gauge bosons are postulated
as the $SU(3)$ symmetric scalar fields.
Four scalar field interactions are parameterized by the typical potential
\begin{equation}
\label{higs} V_e (\phi) = V_0 + \mu^2 \phi^2 + \lambda \phi^4
\end{equation}
where $ \mu^2 < 0$ and $\lambda > 0$ are demanded for spontaneous
symmetry breaking. The first term of the right-hand side
corresponds to the bare vacuum energy density representing the zero-point energy.
The vacuum field $\phi$ is shifted by an invariant quantity $\langle
\phi \rangle$, which satisfies
\begin{equation}
\label{higs1}
\langle \phi \rangle^2 = \phi_0^2 + \phi_1^2 + \cdot \cdot + \phi_i^2
\end{equation}
where the condensation of pseudo-scalar fields is $\langle \phi
\rangle = (\frac{- \mu^2}{2 \lambda})^{1/2}$. DSSB is relevant
for the surface term in (\ref{grac}), $\Theta \frac{g_i^2}{16
\pi^2} Tr F^{\mu \nu} \tilde F_{\mu \nu}$, which explicitly
breaks down the $SU(3)_I$ gauge symmetry to the $SU(2)_L \times
U(1)_Y$ gauge symmetry through the condensation of pseudo-scalar
bosons; scalar fields are also broken from the $SU(3)$ to $SU(2)
\times U(1)$ symmetry in this case. $\Theta$ can be assigned
to a dynamic parameter by
\begin{equation}
\label{thev}
\Theta = 10^{-61} \ \rho_G /\rho_m
\end{equation}
with the matter energy density $\rho_m$ and the vacuum energy
density $\rho_G = M_G^4$. The detail of the $\Theta$ constant
will be discussed in Sect. \ref{sec6}. As the system expands,
pseudo-scalar bosons condense and accordingly gauge boson masses
are reduced.

For QWD, being a $SU(3)_I$
gauge theory, there are nine weak gauge bosons ($n_i^2 = 3^2 = 9$), which consist of
one singlet gauge boson $G_0$ with $i=0$, three degenerate gauge bosons $G_1 \sim G_3$
with $i=1$, and five degenerate gauge bosons $G_4 \sim G_8$ with $i=2$
as shown in (\ref{suth}).
In the case of isospin $1$ gauge bosons, $G_3$ has the third component $0$ and
$G_1$ and $G_2$ have the third component $1$.
In the case of isospin $2$ gauge bosons, $G_8$ has the third component $0$,
$G_6$ and $G_7$ have the third component $1$, and
$G_4$ and $G_5$ have the third component $2$.
For the GWS model with the $SU(2)_L \times U(1)_Y$ gauge theory,
one singlet gauge boson $G_0$ with
$i=0$, three gauge bosons $G_1 \sim G_3$ with $i=1$, and one gauge boson $G_8$ with
$i=2$ are required.
This implies that the mixing of $G_3$ with $i=1$ and $G_8$ with $i=2$ is
the same as mixing between the third component $0$ gauge bosons, which is represented
by the Weinberg mixing angle $\sin^2 \theta_W \simeq 1/4$.
In the DSSB mechanism of the $SU(3)_I$ to $SU(2)_L \times U(1)_Y$ gauge symmetry,
more massive gauge bosons with $i=2$ reduce their roles as intermediate vector bosons.
Even though their contributions are very weak, they appear in quark flavor mixing.

The concept of $SU(2)$ isospin degree of freedom introduces intrinsic quantum numbers.
This means that the intrinsic isospin degree of freedom can be regarded as an intrinsic angular momentum
such as the spin degree of freedom.
The isospin principal number $n_i$ in intrinsic space quantization is very much analogous to the
principal number $n$ in extrinsic space quantization and the intrinsic angular momenta
are analogous to the extrinsic angular momentum.
The total angular momentum has the form of
\begin{math}
\vec J = \vec L + \vec S + \vec I ,
\end{math}
which is the extension of the conventional total angular momentum $\vec J = \vec L + \vec S$.
The intrinsic principal number $n_i$ denotes the intrinsic spatial dimension
or radial quantization: $n_i = 3$ represents weak interactions as an $SU(3)_I$ gauge
theory.

\subsection{Effective Coupling Constant and Gauge Boson Mass}

Weak interactions are generated by the emission and absorption of weak gauge bosons.
Gauge bosons are the analogs of photons for the electromagnetic force and gluons for the color
force. Contrary to the photon, the gauge boson must be massive, otherwise it would
have been directly produced in weak decays. $G_F$ is replaced gauge boson propagation
$\sqrt{2} i_f g_i^2 / 8 (k^2 - M_G^2 )$ with the gauge boson mass $M_G$ and, in
contrast to the dimensionless coupling constant $g_i$ and the isospin factor $i_f$,
has the dimension of inverse energy squared.

The weak interaction amplitude is thus of the form
\begin{equation}
\label{prgr} {\cal M} = - \frac{i_f g_i^2}{4} J^{\mu} \frac{1}
{k^2 - M_G^2}  J_\mu^\dagger = \sqrt{2} G_F J^{\mu}
J_{\mu}^\dagger
\end{equation}
where ${\cal M}$ is the product of two universal current densities and $i_f$ is the isospin factor,
which is defined by
\begin{equation}
\label{isof}
i_f = \frac{1}{4} (i_3^\dagger \lambda^a i_1) (i_2^\dagger \lambda_a i_4)
\end{equation}
with the isospin fields, $i_i$ with $i=1 \sim 4$, in analogy with the color factor $c_f$ in QCD.
The (V - A) current is conserved but the (V+A) current is not conserved.
If $k^2 << M_G^2$, the effective coupling constant becomes
\begin{equation}
\label{glpr}
\frac{G_F}{\sqrt{2}} = - \frac{i_f g_i^2}{8 (k^2 - M_G^2 )} \simeq \frac{g_w^2}{8 M_G^2}
\end{equation}
with $g_w = i_f g_i$ and the weak currents interact essentially at a point.
That is, in the low momentum transfer, the propagation between the currents disappears.
The above equation prompts the idea that weak interactions are weak not
because $g_i$ is much smaller than $e$ but because $M_G^2$ is large.
Indeed, the two coupling constants are related by $e^2 = i_f g_i^2 = g_i^2/16$ and
around energies of the intermediate vector bosons, weak interactions would become of
a strength comparable to the electroweak interactions.

The Lagrangian density as an $SU(2)_L \times U(1)_Y$ gauge theory has the same
form as the one for QWD as an $SU(3)_I$ gauge theory:
\begin{eqnarray}
{\cal L}_{GWS} & = & - \frac{1}{2} Tr  G_{\mu \nu} G^{\mu \nu} +
\sum_{i=1}  \bar \psi_i i \gamma^\mu D_\mu \psi_i \nonumber \\
& + & \Theta \frac{g_w^2}{16 \pi^2} Tr G^{\mu \nu} \tilde G_{\mu
\nu},
\end{eqnarray}
where the bare $\Theta$ term \cite{Hoof2} is a non-perturbative term added to the
perturbative Lagrangian density with $SU(2)_L \times U(1)_Y$ gauge invariance.
The above Lagrangian density has the same form as the GWS model in the fermion and gauge boson parts but
the $\Theta$ term replaced with the Higgs term in the GWS model.
The $\Theta$ term is apparently odd under the P, T, C, and CP operations. The
coupling constant $g_w^2 = i_f g_i^2 = \sin^2 \theta_W g_i^2 = g_i^2/4$ is given in
terms of the weak coupling constant $g_i$ and the isospin factor $i_f$.

Since the covariant derivative is changed from
$D_\mu = \partial_\mu + i g_iG_{\mu a} \lambda^a/2$ in the $SU(3)_I$ gauge theory
to $D_\mu = \partial_\mu + i g_w W_{\mu a} \lambda^a/2 + i g_y B_{\mu}$ in the above Lagrangian density,
the gauge boson mass term is obtained via
\begin{eqnarray}
\triangle {\cal L} & = & \frac{1}{2} (D_\mu \phi)^2 - \frac{1}{2} g_w^2 \langle \phi
\rangle^2 W_\mu W^\mu \nonumber \\ & = & \frac{g_w^2}{2} (W_\mu \phi)^2 - \frac{1}{2}
g_w^2 \langle \phi \rangle^2 W_\mu W^\mu  \cdot \cdot \cdot
\end{eqnarray}
where the intermediate vector bosons $W_\mu$ and $B_\mu$ are defined in the following
section and $\langle \phi \rangle$ is the condensation of the pseudo-scalar boson.
Note that the vacuum energy due to the scalar boson $\phi$ is shifted
with respect to its condensation $\langle \phi \rangle$; this implies that the condensation
subtracts the zero-point energy in the system. The coupling constant $i_f g_i^2$ and
the vacuum expectation value $\langle \phi \rangle$
for the condensation of the pseudo-scalar field make the gauge boson at low energy less massive:
the gauge boson mass is generally defined by
\begin{equation}
\label{grms} M_G^2 = M_H^2 - i_f g_i^2 \langle \phi \rangle^2 = i_f g_i^2
[\phi^2 - \langle \phi \rangle^2]
\end{equation}
where $M_H = \sqrt{i_f} g_i \phi$ indicates the unification gauge boson mass at a
phase transition just above the weak phase transition, $\phi$ denotes the real
scalar boson and $\langle \phi \rangle$ stands for the condensation of the pseudo-
scalar boson. Note that the isospin factor $i_f$ used in (\ref{grms}) is the
symmetric factor for a gauge boson with even parity and the asymmetric factor for a
gauge boson with odd parity. This process leads to the breaking of discrete symmetries P,
C, T, and CP, as is discussed in the following subsection. The crucial point is
that the more massive gauge boson at higher energies becomes lighter through its
condensation at low energies. The Fermi weak interaction constant at low momentum transfer can be, by
the expression (\ref{glpr}), related to the gauge boson mass. The mass must be
identical to the inverse of the screening length, that is, $M_G = 1/ l_{EW} \simeq
G_F^{1/2}$. The gauge boson mass $M_G$ is related to the effective vacuum energy
density $V_e (\phi)$ in (\ref{higs}) by $V_e = M_G^4$: $V_0 = M_H^4 \approx
10^{12} \ \textup{GeV}^4, \ \mu^2 = -2 i_f g_i^2 M_H^2 \approx - 3 \times 10^6 \
\textup{GeV}^2, \ \lambda = i_f^2 g_i^4 \approx 0.01$ for $M_G \approx 10^2$ GeV, $M_H
\approx 10^3$ GeV, and $\alpha_i \approx 0.12$. The weak vacuum represented by massive
gauge bosons is quantized by the maximum wavevector mode $N_R \approx 10^{30}$, the
total gauge boson number $N_G = 4 \pi N_R^3/3 \approx 10^{91}$, and the gauge boson
number density $n_G = \Lambda_{EW}^3 \approx 10^{47} \ \textup{cm}^{-3}$.
The Yukawa potential associated with the massive gauge boson is given by
\begin{equation}
\label{yumu}
V (r) = \sqrt{\frac{i_f g_i^2}{4 \pi}} \frac{e^{-M_G (r - l_{EW})}}{r}
\end{equation}
which shows the short range interaction.

\subsection{Breaking of Discrete Symmetries}

The normal vacuum proceeds to the physical vacuum through the condensation mechanism
represented by DSSB. DSSB by the condensation of scalar fields, which are postulated as spatially longitudinal
components of the gauge bosons, leads to
the breaking of unitarity symmetry as well as the breaking of discrete symmetries.
Electromagnetic and weak interactions preserve different discrete symmetries. The
electromagnetic interaction violates isospin symmetry but preserves the C, P, and T
symmetries. The weak interaction separately violates all of them but is invariant
under the combined CPT symmetry.

In the above, it is briefly mentioned that the $SU(2)_L \times U(1)_Y$ gauge theory
for electroweak interactions has its origin in the $SU(3)_I$ gauge theory for weak
interactions as shown by the triggering of DSSB by the condensation of the scalar bosons.
The condensation is also the origin of the discrete symmetry violation which is
observed in electroweak interactions. The discrete symmetries P, C and T are broken down
explicitly by the condensation of the scalar bosons; the product symmetry CPT
remains intact and CP conserves approximately. Since the condensation of scalar
boson is relevant for the vector and axial-vector currents, the (V - A)
doublet current is conserved in the $SU(2)_L$ weak theory, $\partial_\mu J_\mu^{1 -
\gamma^5} = 0$, but the (V + A) singlet current is not conserved,
\begin{equation}
\partial_\mu J_\mu^{1 + \gamma^5} = \frac{N_f g_i^2}{16 \pi^2} Tr F^{\mu \nu} \tilde F_{\mu \nu}
\end{equation}
with the flavor number $N_f$. The reason for the parity nonconservation is the
scalar boson condensation or the fermion condensation represented by the condensation
number $N_{sc}$ in the fermion mass generation as discussed later. The $SU(3)_I$ symmetry is
broken by DSSB, in which the right-handed symmetry is not manifested in the particle
spectrum such as the absence of right-handed neutrinos explained by P violation
\cite{Lee}. The Cooper pairing between matter particles violates $C$ symmetry and the
$\Theta$ vacuum violates $T$ symmetry explicitly. The inclusion of the third quark
generation is represented by the Kobayashi-Maskawa (KM) matrix \cite{Koba} which
contains P violation and CP symmetry violation. The assumption of the KM matrix has a
verifiable consequence for the decay of the K mesons. CP violation observed in the
neutral kaon decay \cite{Chri} with the probable value $\Theta \simeq 10^{-3}$
indicates T violation because CPT symmetry is conserved. The electric dipole moment of
electrons is observed to be $d_e = -1.5 \times 10^{-26}$ e cm \cite{Murt}, which implies
$\Theta \simeq 10^{-4}$ if the effective dipole length of electrons $l_e \simeq G_F
m_e \simeq 10^{-22}$ cm is used. A possible lepton-antilepton asymmetry is suggested
as a consequence of C, T, and CP violation during the DSSB of the $SU(3)_I$ symmetry
to the $SU(2)_L \times U(1)_Y$ symmetry: the electron asymmetry $\delta_e \simeq
10^{-7}$ \cite{Roh1,Roh11} is suggested as a consequence of the $U(1)_Y$ gauge theory.
The lepton number is not conserved above the weak scale, but the lepton number is
conserved below the weak scale as illustrated by the $U(1)_Y$ gauge theory in the weak
interactions.

The DSSB of gauge symmetry and chiral symmetry induces the (V+A)
current anomaly represented by the value $\Theta \approx 10^{-4}$
at the weak scale. This implies the reduction of zero modes
through the scalar boson condensation. The $\Theta$ vacuum as the
physical vacuum is achieved from the normal vacuum, which
possesses larger symmetry group than the physical vacuum. The
instanton mechanism, vacuum tunneling is expected in the
Euclidean spacetime. The $\Theta$ vacuum term represents the
surface term since it is total derivative and decreases as the
system expands. The condensation of vacuum decreases the mass of
gauge boson, which causes the expansion of the system during a
phase transition: this is the source of the exponential inflation
as expected by the Higgs mechanism.

Photons as NG bosons in DSSB play the role of massless gauge bosons responsible for the $U(1)_e$ gauge theory.
Photons are generated as massless gauge bosons from the weak interactions by DSSB
whereas gauge bosons as intermediate vector bosons are generated as massive gauge bosons by DSSB.
Photons mediate the Coulomb interaction
in the static limit as a result of the symmetry breaking of the $SU(2)_L \times U(1)_Y$ to $U(1)_e$ gauge symmetry.
Detail of photon dynamics generation is described in the following section.

\section{Generation of Electroweak Interactions}
\label{sec3}

In the previous section, the common features in DSSB are addressed, and in this section
the precise generation of the GWS model from QWD is more in particular focused on and the
resolution for problems of the GWS model is suggested.

The generation of the GWS model by $SU(2)_L \times U(1)_Y \rightarrow U(1)_e$ electroweak
interactions \cite{Glas} from QWD as an $SU(3)_I$ gauge theory will now be considered.
Spatially longitudinal components of the gauge bosons parameterized by scalar fields
play the role of Higgs bosons in the electroweak
interactions, so that the GWS model, $SU(2)_L \times U(1)_Y$ gauge
theory, dynamically results from QWD being an $SU(3)_I$ gauge theory.  QWD, an
$SU(3)_I$ gauge theory, generates the electroweak theory, an $SU(2)_L \times U(1)_Y$
gauge theory, through the condensation of the scalar field. The
condensation of the scalar boson also produces DSSB yielding the $U(1)_e$ gauge
symmetry as expected by the Higgs mechanism of electroweak interactions. The covariant
derivative $D_\mu =
\partial_\mu + i g_i G_{\mu a} \lambda^a /2$ of the $SU(3)_I$
gauge theory becomes $D_\mu = \partial_\mu + i g_w W_{\mu a}
\lambda^a /2 + i g_y B_\mu$ of the $SU(2)_L \times U(1)_Y$ gauge
theory. The weak coupling constant $g_i$ is the unification of the
electroweak charge $g_w$ and the hypercharge $g_y$ at the weak scale.

There are several supporting clues for the generation of the electroweak interactions
being an $SU(2)_L \times U(1)_Y$ gauge theory from the weak interactions as an $SU(3)_I$ gauge
theory: parity violation, the Fermi weak coupling constant, the Weinberg angle, three
generations as doublets of leptons and quarks, the Cabbibo angle, and the weak
coupling constant $g_w$. Since scalar field generators do commute with some octet
gauge bosons, massive gauge bosons reduce their masses through the condensation of
pseudo-scalar bosons. This is exactly expected from electroweak interactions
of the $SU(2)_L \times U(1)_Y$ gauge symmetry since three intermediate vector bosons,
$W^{\pm}$ and $Z$ (or $W^{3}$), associated with the generators $\lambda_1 \sim
\lambda_3$ are massive and the photon is massless. Through the condensation of
pseudo-scalar bosons, DSSB from the $SU(2)_L \times U (1)_Y$ to the $U (1)_{e}$ is
accomplished.  The weak gauge bosons $G_4 \sim G_7$ with isospin two are heavier than weak
gauge bosons $G_1 \sim G_3$ with isospin one and their contribution almost disappears
during the phase transition of $SU(3)_I$ to $SU(2)_L \times U(1)_Y$ symmetry: the
detailed concept of isospin will be discussed on treating intrinsic quantum
numbers. During DSSB, parity violation and charge conjugation violation are maximal
due to the condensation of pseudo-scalar fields. The linear combination of the
diagonal isospin octet generators in two octets of triplet isospin combinations
generates the three families of an isospin doublet in leptons and quarks. The mass generation
scheme for fermions is suggested in terms of the DSSB of chiral symmetry without
introducing new input parameters, unlike the GWS model. In this scheme, the $SU(2)_L
\times U(1)_Y$ symmetry and the $U(1)_e$ symmetry using the symmetric isospin factors,
$i^s_f = (i_f^z, i_f^w, i_f^y, c_f^e) = (1/3, 1/4, 1/12, 1/16)$, are applied to the
typical electroweak interactions.

The conclusive clues are presented in the following:
the Fermi weak coupling constant, the Weinberg angle and neutral current, the three generations
of leptons and quarks, the electroweak coupling constants, and the
electroweak flavor mixing angle for quarks.

\subsection{Fermi Weak Coupling Constant}

As described by the previous section, the electroweak interaction amplitude takes the form of
\begin{equation}
\label{prgr1}
{\cal M} = - \frac{g_w^2}{2} J^{\mu} \frac{1} {k^2 - M_W^2}  J_\mu^\dagger
= \frac{4}{\sqrt{2}} G_F J^{\mu} J_{\mu}^\dagger
\end{equation}
where the Fermi weak coupling constant
\begin{equation}
\label{feco}
\frac{G_F}{\sqrt{2}} = - \frac{g_w^2}{8 (k^2 - M_W^2)}  \simeq \frac{g_w^2}{8 M_W^2}
\end{equation}
has the dimension of inverse energy squared.
The isospin factor $i_f = \frac{1}{4} (i_3^\dagger \lambda^a i_1) (i_2^\dagger \lambda_a i_4)$ is defined
in terms of the isospin field $i$.
Note that (\ref{feco}) is exactly identical to (\ref{glpr}), since $g_z = i_f g_i= g_i/3$, $g_w = g_z \cos \theta_W$,
$M_G^2 = M_W^2/i_f$, and $M_Z = M_W / \cos \theta_W$:
the isospin factor $i_f = 1/3$ for the symmetric sextet configuration is discussed in the following subsections.
The gauge boson mass $M_G$ is connected with the electroweak cutoff scale $\Lambda_{EW}$.

\subsection{Weinberg Angle and Neutral Current}

Another clue for the generation of the $SU(2)_L \times U(1)_Y$ to the $SU(3)_I$
symmetry comes from the Weinberg angle $\theta_W$ which represents the mixing of $G_3$
and $G_8$. The experimental value $\sin^2 \theta_W \approx 0.234$ \cite{Kim} is very
close to the theoretical value $\sin^2 \theta_W = g_y^2/(g_w^2 + g_y^2) = 0.25$ if the
coupling constant $g_w$ for the $SU(2)_L$ group and the coupling constant $g_y$ for
the $U (1)_Y$ group are determined from the $SU(3)_I$ group; the theoretical value of
$tan \theta_W = g_y /g_w$ is just the ratio, $1/\sqrt{3}$, between the gauge fields
$G_3$ and $G_8$ from the $SU(3)_I$ gauge symmetry. The neutral weak coupling constant
$g_z = \sqrt{g_w^2 + g_y^2}$ is defined and accordingly $\sin \theta_W = g_y / g_z$
and $\cos \theta_W = g_w/g_z$; $g_w$ and $g_y$ are related by a number known as the
isospin factor of the coupling constants. Notice that the experimental value
$0.236$ is much closer to the theoretical prediction $\sin^2 \theta_W = 0.25$ of this
scheme than $\sin^2 \theta_W = 3/8$ of the $SU(5)$ gauge theory \cite{Geor} at the
tree level.

The Weinberg angle is closely related to massive gauge boson and
massless photon generation. The gauge mass terms come from
equation (\ref{grms}), evaluated at the shifted vacuum $\phi^{'2}
= \phi^2 - \langle \phi \rangle^2$ with scalar boson $\phi$ and condensed scalar boson
$\langle \phi \rangle$. The relevant terms after the phase transition of the
$SU(3)_I$ symmetry to the $SU(2)_L \times U(1)_Y$ symmetry are
\begin{eqnarray}
&&\phi^{'2} (g_i G_\mu^a \lambda^a ) (g_i G^{\mu b} \lambda^b
) \nonumber \\ & \rightarrow & \phi^{'2} [g_w^2 (G_\mu^1)^2 +
g_w^2 (G_\mu^2)^2 + (- g_w G_\mu^3 + g_y G_\mu^8)^2 ] . \nonumber
\end{eqnarray}
Recall that $G_\mu^1 \sim G_\mu^3$ and $G_\mu^8$ are respectively equivalent to $W_\mu^1 \sim W_\mu^3$ and $B_\mu$ in the GWS electroweak model.
There are massive bosons
\begin{eqnarray}
W_\mu^{\pm} & = & \frac{1}{\sqrt{2}} ( G_\mu^1 \mp i G_\mu^2) , \\
Z_\mu^{0} & = & \cos \theta_W G_\mu^3 - \sin \theta_W G_\mu^8
\end{eqnarray}
where their masses are
\begin{eqnarray}
M_W^2 & = & g_w^2 (\phi^{2} - \langle \phi \rangle^2) , \nonumber \\
M_Z^2 & = & M_W^2/ \cos^2 \theta_W = i_f g_i^2 [\phi^2 - \langle
\phi \rangle^2] = \frac{1}{3} g_i^2 [\phi^2 - \langle \phi
\rangle^2] \nonumber
\end{eqnarray}
with $M_H = i_f g_i^2 \phi^2$ respectively. The fourth vector
identified as the photon, orthogonal to $Z^0$, remains massless:
\begin{equation}
\label{phot}
A_\mu = \sin \theta_W G_\mu^3 + \cos \theta_W G_\mu^8
\end{equation}
with the mass $M_A = 0$.
Photons play the role of massless gauge bosons responsible for the $U(1)_e$ gauge theory.
They are understood as NG bosons during DSSB \cite{Namb}.
Two physical neutral gauge fields $Z_\mu$ and $A_\mu$ are orthogonal
combinations of the gauge fields $G_\mu^3$ and $G_\mu^8$ with the mixing angle $\theta_W$.
The mixing represents the mixing of gauge bosons with the zero third component of isospin,
$G_\mu^3$ with isospin $1$ and $G_\mu^8$ with isospin $2$.
The electromagnetic current is the combination of the two neutral
currents $J_\mu^3$ and $j_\mu^y$.
The generators of this scheme satisfy $\hat Q = \hat I_3 + \hat Y/2$ as shown in (\ref{chqu}) so that
\begin{equation}
j_\mu^e = J_\mu^3 + j_\mu^y/2 .
\end{equation}
The hypercharge operator $\hat Y = \hat B - \hat L$ is used and applied in the subsection of the
three generations of leptons and quarks.
The interaction in the neutral current sector can be given by
\begin{eqnarray}
\label{necu} && -i g_w J_\mu^3 G^{3 \mu} - i g_y j_\mu^y G^{8 \mu} /2 \nonumber \\
& = & - i e j_\mu^e A^\mu - i g_z [J_\mu^3 - \sin^2 \theta_W j_\mu^e] Z^\mu
\end{eqnarray}
where the relation
\begin{equation}
e = g_w \sin \theta_W = g_y \cos \theta_W = g_z \cos \theta_W \sin \theta_W
\end{equation}
is used.
The Weinberg angle in the neutral current interaction of (\ref{necu}) is further addressed by
\begin{eqnarray}
- i \frac{g_w}{\cos \theta_W} j_\mu^{NC} Z^\mu
& = & - i g_z \bar \psi \gamma^\mu
[\frac{1}{2} (1 - \gamma^5 ) I_3 - \sin^2 \theta_W Q] \psi Z_\mu
\nonumber \\ & = & - i g_z \bar \psi \gamma^\mu \frac{1}{2} (g_V -
g_A \gamma^5) \psi Z_\mu
\end{eqnarray}
where $j_\mu^e$ is the electric current density.
The vector and axial-vector couplings, $g_V$ and $g_A$, are determined by
\begin{equation}
g_V = I_3 - 2 \sin^2 \theta_W Q, \ \ g_A = I_3 .
\end{equation}
Experimental values for the vector and axial-vector coupling from
neutrino-electron data \cite{Hung} are reasonably in good
agreement with isospin quantum numbers shown in Table \ref{isqu},
which is obtained by the GWS model, when the Weinberg angle is
$\sin \theta_W = 1/2$; for the electron, the experimental values are
$g_A = - 0.52$ and $g_V = 0.06$, and theoretical values are $g_A =
- 0.5$ and $g_V = 0$.

\subsection{Electroweak Coupling Constants}

During the phase transition from the $SU(3)_I$ gauge theory to the $SU(2)_L \times U(1)_Y$ gauge theory,
two fermion interactions with triplet isospins are represented by $3 \otimes 3 = \bar 3 \oplus 6$ in group theoretical language.
Similarly to the strong interactions, the triplet isospin charges $(A, B, C)$ are introduced.
From two fermions, a triplet which are asymmetric combinations becomes
\[
(AB - BA)/\sqrt{2}, \ (BC - CB)/\sqrt{2}, \ (CA - AC)/\sqrt{2}
\]
and a sextet of symmetric combinations becomes
\begin{eqnarray}
&& AA, \ BB, \ CC, \nonumber \\
(AB + BA)/\sqrt{2}, \ && (BC + CB)/\sqrt{2}, \ (CA + AC)/\sqrt{2}
. \nonumber
\end{eqnarray}
The isospin factor $i_f = - 2/3$ for the triplet configuration is obtained for the fermion-fermion interactions.
Since six pairs with the same isospin factor are, together with the
normalization factor $1/6$, taken into account, the isospin factor is $i_f = 1/3$.
The completely symmetric sextet configuration is related to the three generations of leptons and quarks;
sextet members might lead to three generations of the $SU(2)_L$ isospin symmetry after the electromagnetic phase transition:
$SU(2)_L \times U(1)_Y \rightarrow U(1)_e$.
The involved gauge bosons in the $SU(2)_L \times U(1)_Y $ gauge theory are $G_1 \sim G_3$ and $G_8$.
Note that the coupling constant for the charge neutral current is
$\alpha_z = i_f \alpha_i = \alpha_i/3$.
Leptons are asymmetric in spin and isospin states while quarks are symmetric in spin and isospin states.

This scheme explains the weak coupling constant consistent with
experimental values through the data of muon decay \cite{Bard}.
The observed muon lifetime and mass give a Fermi weak coupling
\begin{equation}
G_F = \frac{\sqrt{2}}{8}(\frac{g_w}{M_W})^2 = 1.166 \times 10 ^{-5} \ \textup{GeV}^{- 2} .
\end{equation}
The corresponding value of the weak coupling constant $g_w$ is given by
$g_w \approx 0.61$ and the weak fine structure constant becomes
\begin{equation}
\alpha_w = g_w^2 /4 \pi \approx 0.03 .
\end{equation}
On the other hand, since $\alpha_e = 1/137$, $\alpha_e = \alpha_w \ \sin^2 \theta_W$, $\alpha_z \cos^2 \theta_W = \alpha_w$,
$\alpha_z = i_f \alpha_i$, and $\sin^2 \theta_W = 1/4$, the prediction of $\alpha_z = g_z^2 /4 \pi = 0.04$
and $\alpha_i = g_i^2/4 \pi \simeq 0.12$ at the weak scale is obtained.
In summary, the coupling constant hierarchy in the weak interactions is
$\alpha_i, \ \alpha_z = \alpha_i/3 \simeq 0.04, \ \alpha_w = \alpha_i/4 \simeq 0.03, \ \alpha_y = \alpha_i/12 \simeq 0.01$, and $\alpha_e = \alpha_i/16 \simeq 1/133$ at the weak scale.
The isospin factor $i_f = 1/4$ for $\alpha_w = \alpha_i/4$ represents the coupling between parallel $SU(2)$ isospins.

\subsection{Three Generations of Leptons and Quarks: Weak Isospin and Charge Quantum Number}

This scheme exhibits three generations of leptons and quarks holding
family symmetry, which is one of the fundamental question in particle physics.
The left-handed lepton doublets are
\begin{equation}
{\nu_e \choose e}_L, {\nu_\mu \choose \mu}_L, {\nu_\tau \choose \tau}_L
\end{equation}
and the right-handed lepton singlets are
$e_R, \mu_R, \tau_R$.
The left-handed quark doublets are
\begin{equation}
{u \choose d}_L, {c \choose s}_L, {t \choose b}_L
\end{equation}
and the right-handed quark singlets are $u_R$ or $d_R$, $c_R$ or $s_R$, and $t_R$ or
$b_R$. The detailed scenarios of the isospin and charge quantum numbers are as follows.

The $SU(3)_I$ symmetry represents, in group theoretical language,
$3 \otimes 3 \otimes 3 = 10 \oplus 8 \oplus 8 \oplus 1$ for three
fermion combinations with triplet isospins. When triplet isospin
charges $A, B, C$ are introduced, the two octets are partially
asymmetric under the interchange of isospin. One octet is asymmetric
under the interchange of the first and second isospins, $(A,B)$, and
the other octet is asymmetric under the interchange of the first and
third isospins, $(A,C)$, or the second and third isospins,
$(B,C)$. The linear combination of two octets yields the fermion
families with symmetric isospin configurations. Three families of
leptons are created as the linear combination of diagonal isospin
components of three $SU(2)$ subgroups, I, U, and V isospins, in
two octets. The phase transition of the $SU(3)_I \rightarrow
SU(2)_L \times U(1)_Y \rightarrow U(1)_e$ gauge theory provides us with
the electric charge quantization $\hat Q = \hat I_3 + \hat Y/2$
with $\hat Y = \hat B - \hat L$. For example, the linear
combination of generators $\lambda^a_{3}$ and $\lambda^s_{3}$ is
for left-handed lepton doublet, and the linear combination of
generators $\lambda^a_{8}$ and $\lambda^s_{3}$ is for the right-handed
lepton singlet, where the superscripts $a$ and $s$ represent the
asymmetric and symmetric states under interchanging the first two
particles, respectively. The generators are dual in parity:
\begin{eqnarray}
\lambda^a_3/2 & = & diag (1,-1,0)/2, \ \lambda^s_3/2 =
diag(1,1,0)/2, \nonumber \\
\lambda^a_8/2 \sqrt{3} & = & diag (1,1,-2)/6, \lambda^s_8/2
\sqrt{3} = diag (1,1,2)/6. \nonumber
\end{eqnarray}
Note that the rotation by the Weinberg angle is included in the
coefficients. Three generations of quarks are also created as a
linear combination of the diagonal isospin components of I, U and V
subgroups: for instance, the linear combination of
$\lambda^a_{3}$ and $\lambda^s_{8}$ for left-handed quark doublet
and the linear combination of $\lambda^a_{8}$ and $\lambda^s_{8}$
for the right-handed quark singlet. The mass difference among the three
families depends on the condensation of the pseudo-scalar bosons. The
$SU(3)_I$ gauge symmetry is broken as a result of the
condensation of the pseudo scalar bosons. The $\lambda^s_{3}$ is
relevant for lepton families with $I_3 = 0$ and $Y = - L = -1$,
the $\lambda^s_{8}$ is relevant for the quark families with $I_3 = 0$
and $Y = B = 1/3$, the $\lambda^a_{3}$ is relevant for the
left-handed doublets $I_3 = \pm 1/2$ and $Y = 0$, and the
$\lambda^a_{8}$ is relevant for the right-handed singlets with $I_3 =
0$ and $Y = -1$. Note that the states of $\lambda^a_{3}$ and
$\lambda^a_{8}$ determine the distinction between the left- and
right-handed particles; they are closely related to the axial-vector
charges since they hold odd parity while $\lambda^s_{3}$ and
$\lambda^s_{8}$ are related to the vector charges since they hold
even parity. A summaries for the generation of leptons and
quarks is as follows. The linear combination of $\lambda^a_{3}$
and $\lambda^s_{3}$ generates the left-handed lepton doublets with
$I_3 = \pm 1/2$ and $Y = - L = -1$: $diag(1,-1,0)/2 -
diag(1,1,0)/2 = diag(0,-1,0)$. The linear combination of
$\lambda^a_{3}$ and $\lambda^s_{8}$ generates the left-handed quark
doublets  with $I_3 = \pm 1/2$ and $Y = B = 1/3$: $diag(1,-1,0)/2
+ diag(1,1,2)/6 = diag(2,-1,1)/3$. The linear combination of
$\lambda^a_{8}$ and $\lambda^s_{3}$ generates the right-handed lepton
singlets with $I_3 = 0$ and $Y = - 2$: $- diag(1,1,-2)/6 -
diag(1,1,0)/2 = diag(-2,-2,1)/3$. The linear combination of
$\lambda^a_{8}$ and $\lambda^s_{8}$ generates the right-handed quark
singlets with $I_3 = 0$ and $Y = 4/3$ or with $I_3 = 0$ and $Y = -
2/3$: $- diag(1,1,-2)/6 + diag(1,1,2)/6 = diag(0,0,2)/3$ or $-
diag(1,1,-2)/6 - diag(1,1,2)/6 = diag(-1,-1,0)/3$ where the
element $-1/3$ denotes the down quark and the element $2/3$ denotes the
up quark. Weak isospin and electric charge quantum numbers of
leptons and quarks are summarized in Table \ref{isqu}. This
scheme illustrates the left-right symmetry before DSSB; it is similar
to the composite model \cite{Pati0} and the left-right symmetric model of
weak interactions \cite{Pati}.

In other words, each elementary fermion such as the lepton or
quark possesses color and isospin degrees of freedom in addition
to spin degrees of freedom. The lepton is postulated as a color
singlet state while the quark is postulated as a color triplet
state so that it may interact through the color charge exchange.
All leptons and quarks consist of a lepton octet and a quark octet, in
which each family explicitly possesses three generations of
isospin doublets with one electric charge unit difference: this is
very much analogous to the celebrated eight-fold way in hadron
spectra with quark flavor symmetry.

\subsection{Electroweak Flavor Mixing Angle for Quarks}

Further clues are the mixing angles of the KM matrix \cite{Koba} including the Cabbibo angle \cite{Cabb}.
The mixing between the $d$ and $s$ quarks in the decay of the vector boson $W^+$ is
represented by the Cabbibo angle which indicates mixing among quark families.
The mixing of the d and s quark is observed by comparing $\Delta S = 1$ and $\Delta S = 0$ decays:
\begin{eqnarray}
\frac{\Gamma ( K^+ \rightarrow \mu^+ + \nu_\mu)}{\Gamma ( \pi^+
\rightarrow \mu^+ + \nu_\mu)} & \sim & \sin^2 \theta_C, \nonumber
\\ \frac{\Gamma ( K^+ \rightarrow \pi^0 + e^+ + \nu_\mu)}{\Gamma (
\pi^+ \rightarrow \pi^0 + e^+ + \nu_\mu)} & \sim & \sin^2 \theta_C
.
\end{eqnarray}
The mixing between the quarks is more extended to the third generation of quarks by the KM matrix which includes three angles
and one CP phase angle \cite{Koba,Glas1}.
These mixing angles and this phase angle show the possibility that quark
interactions are generated from  both the weak interaction of the $SU(3)_I$ gauge symmetry and
the strong interaction of the $SU(3)_C$ gauge symmetry.

Recall that electric charges for left-handed quarks are obtained
by the linear combination of generators $\lambda^a_{3}$ and
$\lambda^s_{8}$; the electric charge matrix is given by $diag
(2/3, -1/3, 1/3)$. The existence of the third component in QWD,
which has a quantum number $Y = 2/3$, which is unpredictable from the GWS
model, indicates the $SU(3)_I$ flavor symmetry and is related to
the mixing of $W^+ = G^{1 + i2}$ and $G^{4 + i5}$ gauge fields for
positively charged gauge fields; for instance, the total current
becomes $j_\mu = \cos \theta_C \ j_\mu^{1 + i2} + \sin \theta_C \
j_\mu^{4 + i5}$ where the Cabbibo angle $tan^2 \theta_C$ is
determined by the mass square ratio of the two gauge fields $G^{1
+i2}$ and $G^{4 +i5}$: $M_W = M_G^{1 +i2}$ and $M_G^{4 +i5} \simeq
2 M_W$. The Cabbibo angle $\sin \theta_C = (M_G^{1 + i2}/M_G^{4 +
i5})^2 \simeq 1/4$ is thus predicted and is very close to the
experimental value $\sin \theta_C = 0.231$. This is also supported
by $\sin^2 \theta_C$ being connected with the mass ratio
$\approx 1/20$ of the $d$ and $s$ quarks \cite{Wein7,Frit2}. The
Cabbibo angle thus indicates the mixing between the $SU(2)$
subgroups I and U. This can be considered to be conclusive evidence of the massive
gauge bosons $G_4$ and  $G_5$ with the masses $2 M_W$ in addition to
three intermediate vector bosons.

The other two mixing angles in the KM matrix respectively
represent the mixing between the $SU(2)$ subgroups I and V and
the mixing between the subgroups U and V. Since the other mixing
angles are experimentally close to zero, the masses of the gauge
bosons $G_6$ and $G_7$ must be heavier than those of $G_1$ and
$G_2$. Note that weak CP violation from the decay of the neutral
kaon is relevant for the $\Theta \leq 10^{-3}$, and the absence of
right-handed neutrinos is a direct consequence of the
non-conservation of the (V + A) current under the breaking of
discrete symmetries.

\section{Grand Unification of Strong and Electroweak Interactions}
\label{sec4}

In this section, the grand unification of QWD and QCD is clarified in terms of
the experimental strong and weak coupling constants.
A coupling constant hierarchy for weak interactions is suggested in analogy with
coupling constant hierarchy for the strong interactions.

A grand unified group $H$ contains the $SU(3)_I$ group for the weak interactions
and the $SU(3)_C$ group for strong interactions \cite{Frit,Roh3,Roh31} as
subgroups: $H \supset SU(3)_I \times SU(3)_C$.
A certain grand unification group $H$ breaks down to the $SU(3)_I \times
SU(3)_C$ group at the grand unification scale around $10^{3}$ GeV,
which is much lower than the grand unification scale $10^{15}$ GeV of the GUT \cite{Geor}
being the grand unification of the standard model $SU(3)_C \times SU(2)_L \times U(1)_Y$.
This scheme might provide the resolution to the hierarchy problem and the
analogy between the $SU(3)_I$ and $SU(3)_C$ symmetries.
In the following, grand unification of the weak and strong interactions and the comparison of
effective coupling constants are discussed.

\subsection{Grand Unification of Quantum Weakdynamics and Quantum Chromodynamics}

There is conclusive evidence for the proposal that QCD for the strong force is unified with QWD for the weak force
at the grand unification scale around $10^{3}$ GeV.
The fine structure constant $\alpha_s$ for the strong interactions is measured by
numerous experiments \cite{Hinc}:
\begin{equation}
\label{stco}
\alpha_s (M_Z) \simeq 0.12
\end{equation}
which has been evaluated at the momentum of the $Z$ boson mass $q = M_Z$.
The fine structure constant $\alpha_w$ for the $SU(2)_L$ weak
interaction is given by the data of muon decay \cite{Bard}:
\begin{equation}
\label{weco}
\alpha_w (M_W) \simeq 0.03
\end{equation}
at the momentum of the $W$ boson mass $q = M_W$.
The two empirical coupling constants (\ref{stco}) and (\ref{weco}) provide the grand
unification scale around $10^{3}$ GeV, where the strong coupling constant is the same as the weak coupling constant:
$\alpha_h = \alpha_s = \alpha_i$.
For the $SU(3)_I$ symmetry, a symmetric sextet configuration in isospin-isospin
interactions has the fine structure constant $\alpha_z = i_f \alpha_i = \alpha_i/3$, which is the coupling constant
for the gauge boson $Z$.
Since $\alpha_z = \alpha_w/\cos^2 \theta_W$ and $\sin^2 \theta_W = 1/4$, the fine structure constant
for the $W$ boson is $\alpha_w = \alpha_i/4$;
the fine structure constant for QWD, $\alpha_i (M_Z) \simeq 0.12$, is thus obtained.
This leads to the fine structure constant for the electromagnetic interactions
$\alpha_e = \alpha_w \sin^2 \theta_W = \alpha_i /16 = 1/133$.
In summary, the fine structure constants are given by
$\alpha_h = \alpha_s = \alpha_i \simeq 0.12$,
$\alpha_z = \alpha_i /3 \simeq 0.04$, $\alpha_w = \alpha_i /4 \simeq 0.03$, $\alpha_y = \alpha_i /12 \simeq 0.01$,
and $\alpha_e = \alpha_i/16 \simeq 1/133$ around the weak scale $10^2$ GeV.
In addition, the electromagnetic interactions give the fine structure constant $4 \alpha_e/9 = \alpha_i/36$
for up quarks and $\alpha_e/9 = \alpha_i/144$ for the down quarks.
This explains the group chains $H \supset SU(3)_I \times SU(3)_C$ and
$SU(3)_I \supset SU(2)_L \times U(1)_Y \supset U(1)_e$.

After the phase transition of the group $H$, the groups $SU(3)_I$
and $SU(3)_C$ preserve the analogous properties. The isospin
coupling constant is given by $g_i$ and the Weinberg weak mixing angle
is given by $\sin^2 \theta_W = 1/4$.
As the energy scale decreases, the $SU(3)_I$ group for isospin
interactions breaks down to the $SU(2)_L \times U(1)_Y$ group for the
electroweak interactions at the electroweak scale.
The electroweak coupling constants are, in summary,
$\alpha_z = i_f^z \alpha_i = \alpha_i/3 \simeq 0.04$, $\alpha_w =
i_f^w \alpha_i = \alpha_i/4 \simeq 0.03$, $\alpha_y = i_f^y
\alpha_i = \alpha_i/12 \simeq 0.01$, and $\alpha_e = i_f^e
\alpha_i = \alpha_i/16 \simeq 1/133$ for the symmetric isospin
interactions at the weak scale and $-2 \alpha_i/3$, $-
\alpha_i/2$, $- \alpha_i/6$, and $- \alpha_i/8$ for the asymmetric
isospin interactions: $i_f^w = \sin^2 \theta_W$ and $i_f^e =
\sin^4 \theta_W$.  The isospin factors introduced are $i^s_f = (i_f^z,
i_f^w, i_f^y, i_f^e) = (c_f^b, c_f^n, c_f^z, c_f^f) =
(1/3, 1/4, 1/12, 1/16)$ for symmetric interactions and $i^a_f
= (-2/3, -1/2, -1/6, -1/8)$ for asymmetric interactions. The
symmetric charge factors reflect an intrinsic even parity with
repulsive force while the asymmetric charge factors reflect an
intrinsic odd parity with attractive force; this suggests
electromagnetic duality. The symmetric charge factors reflect
intrinsic even parity with repulsive force while the asymmetric
charge factors reflect an intrinsic odd parity with attractive force.
As a consequence of gauge theories,
conservation laws are expected.
Table \ref{coga} shows relations
between conservation laws and gauge theories in the weak interactions.
However, there is a possibility that not for the
separate conservation of the baryon number and the lepton number hold,
but the combined conservation of $(B - L)$ number conservation
in the energies above $10^2$ GeV.

\subsection{Coupling Constants for Fundamental Forces}

The new grand unification energy $10^{3}$ GeV rather than the
conventional energy $10^{15}$ GeV is obtained so that the
hierarchy problem seems to be resolved. The unification at the
order of a TeV energy is consistent with recent GUT \cite{Poma}.
Dynamical symmetrical breaking is adopted instead of the Higgs
mechanism; the condensation of pseudo-scalar bosons reduces the
mass of the gauge boson and it becomes the source of the system
inflation.

QWD has an asymptotic freedom in the weak coupling constant $g_i$ just at
the notable characteristic of QCD is the asymptotic freedom due to
anti-screening at short distance as a non-Abelian gauge theory according to the renormalization group study \cite{Gros}.
Massive gauge bosons at the grand unification scale produce massive intermediate vector bosons and massive gluons
at weak and strong interaction energies respectively.
The effective grand unified coupling becomes
\begin{equation}
\frac{G_H}{\sqrt{2}} = - \frac{g_h^2}{8 (k^2 - M_G^2 )} \simeq \frac{g_h^2}{8 M_G^2} \approx 10^{-4} \ \textup{GeV}^{-2}
\end{equation}
where the coupling constant $g_h$ for the grand unified gauge group $H$:
the effective coupling constant chain becomes $G_H \supset G_F \times G_R$ with
the effective strong coupling constant $G_R /\sqrt{2} = g_s^2/ 8 M_G^2 \approx 10 \ \textup{GeV}^{-2}$.
Note that $G_H$ is close to $G_F$ or it is slightly less than $G_F$, since
$\alpha_h \sim \alpha_i \sim \alpha_s$ at the energy scale $10^3$ GeV.
The mass of the gauge boson at the grand unification scale is expressed by
$M_G^2 = M_H^2$ at a slightly higher phase transition energy above the weak scale.
The gauge boson decreases its mass as the condensation increases when the energy scale decreases.
At $T \simeq 10^2$ GeV, the gauge boson number density is $n_G = M_G^3 \simeq 10^6 \ \textup{GeV}^3 \simeq 10^{47} \ \textup{cm}^{-3}$,
the vacuum energy density is $V_0 = M_G^4 \simeq 10^8 \ \textup{GeV}^4 \simeq 10^{25} \ g \ \textup{cm}^{-3}$, and
the total gauge boson number is $N_G \simeq 10^{91}$, which is a conserved good quantum number.

Of the four fundamental forces in nature, the three forces, the
electromagnetic, strong, and weak forces, are unified at the grand unification scale.
Strong interactions are limited in range to about $10^{-13}$ cm and are insignificant
even at the scale of the atom $10^{-8}$ cm, but play an important role in binding the nucleus.
Weak interactions with an even shorter scale $(\leq 10^{-15} \ \textup{cm})$
do play an important role in weak decay processes.
The strengths of the four forces are roughly in orders of magnitude $10, 10^{-2},
10^{-5}$, and $10^{-41}$ for strong, electromagnetic, weak, and gravitational forces, respectively.
The difference in strength is more than a factor of $G_R/G_N \approx 10^{39}$ for the strong and gravitational interactions.
The ratio of the electroweak and gravitational forces is also obtained by $G_F/G_N \approx 10^{33}$.

The gauge boson possesses the isospin $SU(3)_I$ and color $SU(3)_C$ symmetries below
the grand unification energy. The gauge boson with the energy higher than its mass
thus effectively interacts with the Coulomb potential outside the grand unification
scale while the gauge boson with the energy lower than its mass essentially interacts
with the Yukawa potential within the grand unification or weak scale. At the grand
unification scale, the gauge boson has a mass so that the interaction range is
limited to the grand unification scale. However, as energy goes down the gauge boson
loses mass because of the pseudo-scalar boson condensation. Isospin interactions are
represented by both the Coulomb potential and the Yukawa potential with a Fermi weak
constant $G_F \approx 10^{-5} \ \textup{GeV}^{-2}$. This implies that the isospin
field with the Yukawa interaction does not propagate over the long range due to the
heavy mass; recall that the (V - A) current in the weak theory has a point-like
interaction with a Fermi coupling constant $G_F$. At the electroweak phase transition,
the isospin interaction range for massive gauge bosons is restricted to the weak scale
but the electromagnetic interaction range for the massless photons is infinite. Gauge
boson masses change their values at the different energy scale due to the
scalar boson condensation. For example, $M_G \approx 10^{2}$ GeV at the grand
unification scale or at the electroweak phase transition scale and $M_G \approx 0.1$
GeV at the QCD cutoff scale. The Fermi coupling constant $G_F$ thus denotes the effective
coupling for the Yukawa potential of the isospin field at the electroweak scale.
Energy scales for phase transitions of the fundamental forces are accordingly as follows;
the grand unification corresponds to $E \approx \Lambda_H$, the weak force to $E
\approx \Lambda_W$ with the weak scale $\Lambda_W \approx 10^2$ GeV, and the strong
force to $E \approx \Lambda_{QCD}$ with the strong mass scale $\Lambda_{QCD} \approx
0.1$ GeV.

\section{Quark and Lepton Mass Generation}
\label{sec5}

The DSSB mechanism suggests quark and lepton mass generation, which is the outstanding
problem in the GWS model using the Higgs mechanism. In this scheme, quarks and leptons
are not treated as fundamental elementary particles but as composite particles.  The
duality property before phase transition may be broken by DSSB after phase transition.

Ordinary mass terms in the Lagrangian are not allowed, because the left- and
right-handed components of the various fermion fields have different quantum numbers,
and so simple mass terms violate gauge invariance. To generate masses for the quarks and
leptons, the DSSB mechanism of gauge symmetry and chiral symmetry is required. The
condensation of pseudo-scalar bosons and fermion pairs is connected with the
lepton and quark masses in the course of parity and charge conjugation violation, which
breaks chiral symmetry; the $SU(2)_L$ doublet (V - A) current is conserved but the
$SU(2)_R$ (or $U(1)_R$) singlet (V + A) current is not conserved. The massless gauge
bosons in the DSSB of gauge symmetry and chiral symmetry are photons.

The dual Meissner effect, constituent fermions, fine and hyperfine structure,
and quark and lepton mass generation are addressed in the following.

\subsection{Dual Meissner Effect}

The quark or lepton formation is the consequence of the isospin-isospin interaction
due to the dual Meissner effect, in which the isotope electric monopole and the
isotope magnetic dipole (isospin) are confined inside the quark (or lepton), while the
isotope magnetic monopole and the isotope electric dipole are confined in the vacuum.
The difference number of right- and left-handed (singlet and doublet) fermions $N_{sd} = N_{ss} - N_{sc}$,
the number of left-handed constituent particles $N_{ss}$, and the right-handed condensation number
$N_{sc}$ are introduced.

During the DSSB of gauge symmetry and chiral symmetry, the dual Meissner
effect of the isospin electric field in the relativistic case can be given by
\begin{equation}
\label{wame0}
\partial_\mu \partial^\mu G^\mu = - M_G^2 G^\mu
\end{equation}
where the right-hand side is the screening current, $j^\mu_{sc} = - M_G^2 G^\mu$.
Recall that the masses of the gauge bosons are expressed by
\begin{math}
M_G^2 = M_H^2 - i_f g_i^2 \langle \phi \rangle^2 = i_f g_i^2 [\phi^2 -
\langle \phi \rangle^2]
\end{math}
where $\langle \phi \rangle$ represents the condensation of the
pseudo scalar boson and $i_f$ denotes the isospin factor. The
screening of the isospin field intensity in the isospin
superconducting state is given by
\begin{equation}
\label{meef}
\nabla^2 \vec E_i = M_G^2 \vec E_i
\end{equation}
where $\vec G$ is the isospin electric field $E_i$
excluded in the vacuum by $\vec E_i = \vec E_{i0} e^{- M_G r}$.
Note the difference between the isospin dielectric due to the isospin electric
field $\vec E_i$ and the isospin diamagnetism due
to the isospin magnetic field $\vec B_i$.
The mechanism is by analogy connected with the Faraday induction law
which opposes the change in the isospin electric flux, rather than
the isospin magnetic flux according to Lenz's law.

The gauge boson mass is related to the fermion mass $m_f$:
\begin{equation}
\label{gafe}
M_G = (\frac{g_{im}^2 |\psi (0)|^2}{m_f})^{1/2} \simeq \sqrt{\pi} m_f i_f \alpha_i \sqrt{N_{sd}}
\end{equation}
which is obtained by the analogy of electric superconductivity
\cite{Aitc}, $M^2 = q^2 |\psi(0)|^2/m$: $q= - 2 e$ and $m = 2
m_e$ are replaced by $g_{im}$ and $m_f$. The isospin magnetic
coupling constant $g_{im} = 2 \pi n/\sqrt{i_f} g_i = 2 \pi
\sqrt{N_{sd}}/\sqrt{i_f} g_i$ by the Dirac quantization condition,
is used. $|\psi (0)|^2$ denotes the particle probability density
and $N_{sd}$ denotes the difference number of right- and left-handed
fermions in intrinsic two-space dimensions.

A fermion mass term in the Dirac Lagrangian has the form $m_f \bar
\psi \psi = m_f (\bar \psi_R \psi_L + \bar \psi_L \psi_R)$ where
the mass term is equivalent to a helicity flip. Left-handed
fermions are put into $SU(2)$ doublets and the right-handed ones
into $SU(2)$ singlets. The coherent fermion system is effectively
a collection of the Cooper pairs of left- and right-handed
fermions, so that the macroscopic occupancy of a single quantum
state could occur; all the pairs have the same center of mass
momentum known as the coherent state. The fermion mass generation
mechanism is the dual pairing mechanism of the constituent fermions,
which makes boson-like particles of paired fermions. According to
the electric-magnetic duality \cite{Dira,Mand,Seib}, the isospin
electric flux is quantized by $\Phi_E = \oint \vec E_i \cdot d
\vec A = \sqrt{i_f} g_i$ in the matter space while the isospin
magnetic flux is quantized by $\Phi_B = \oint \vec B_i \cdot d
\vec A = g_{im}$ with the isospin magnetic coupling constant
$g_{im}$ in the vacuum space: the Dirac quantization condition
\begin{equation}
\sqrt{i_f} g_i g_{im} = 2 \pi n = 2 \pi \sqrt{N_{sd}}
\end{equation}
is satisfied with the connection between $n$ and $N_{sd}$. In the matter space, it is
the pairing mechanism of isospin electric monopoles while in the vacuum space, it is
the pairing mechanism of isospin magnetic monopoles according to the duality between
electricity and magnetism \cite{Dira,Mand,Seib}: isospin electric monopole pairing and
isospin magnetic monopole condensation. In the dual pairing mechanism, the discrete
symmetries P, C, T, and CP are dynamically broken. The isospin electric monopole, isospin
magnetic dipole, and isospin electric quadrupole remain in the matter space, but the
isospin magnetic monopole, isospin electric dipole, and isospin magnetic quadrupole
condense in the vacuum space as a consequence of P violation. Antimatter particles
condense in the vacuum space while matter particles remain in the matter space as a
consequence of C violation: the matter-antimatter asymmetry. The lepton-antilepton
asymmetry is also supporting evidence of the discrete symmetry breaking.  The electric
dipole moment of the neutron and the decay of the neutral kaon decay are the typical
examples for T or CP violation.

Normal fermions with the quantum number $N_{sd}$ interact with each
other with isospin symmetric configurations, $c^s_f = (c_f^b,
c_f^n, c_f^z, c_f^f) = (1/3, 1/4, 1/12, 1/16)$, while condensed
fermions with the condensation number $N_{sc}$ interact each other
with isospin asymmetric configurations, $c^a_f = (-2/3, -1/2,
-1/6, -1/8)$. Comparing (\ref{gafe}) with (\ref{grms}), it is
realized that $M_H = \sqrt{\pi} m_f i_f \alpha_i \sqrt{N_{ss}}$
and $\langle \phi \rangle = (m_f^2 i_f \alpha_i N_{sc}/4)^{1/2}$.
Note that equation (\ref{gafe}) is analogous to the fermion mass
$m_f = \langle \bar F F \rangle / 2 \mu^2$ with the condensation
of the technifermion $F$ and the extended technicolor scale $\mu$
in the extended technicolor model \cite{Appe}. It is instructive
that $m_f = \lambda_f M_G$ where $\lambda_f \simeq
\frac{1}{\sqrt{\pi} i_f \alpha_i \sqrt{N_{sd}}}$ depends on the
number $N_{sd}$. The intrinsic quantum number of a constituent fermion is thus important
in determining the fermion masses. For example, the $u$ quark with the
current quark mass $m_u \approx 5$ MeV has $N_{sd}^u = 10^{10}$
and electron with the mass $m_e \simeq 0.5$ MeV has $N_{sd}^e
\simeq 10^{12}$ for the effective coupling constant $G_F = 10^{-5}
\ \textup{GeV}^{-2}$ and the gauge boson mass $M_G \simeq 10^2$
GeV: $m_e/m_u \simeq 1/10$ and $N_{sd}^e/N_{sd}^u \simeq 1/10^2$.
Each quark or lepton holds a different intrinsic quantum number as a distinct state.

\subsection{Constituent Fermions}

Leptons and quarks are postulated to be composite particles consisting of constituent fermions as a result of $U(1)_Y$ gauge theory.
This concept is similar to the concept of technicolors or preons \cite{Suss,Pati} as more fundamental particles forming quarks and leptons.

The relation between the gauge boson mass and the free fermion mass, which is confirmed in (\ref{gafe}), is given by
$M_H = \sqrt{\pi} m_f i_f \alpha_i \sqrt{N_{ss}}$
or
\begin{equation}
\label{gafe5}
M_G = \sqrt{\pi} m_f i_f \alpha_i \sqrt{N_{sd}}
\end{equation}
where $m_f$ is the mass of a fermion, $N_{sd}$ is the difference number of left- and right-handed fermions in intrinsic two-space dimensions,
and $N_{ss}$ is the number of left-handed fermions.
The fermion mass formed as a result of the dual pairing mechanism in the above is composed of constituent particles:
\begin{equation}
\label{gafe1}
m_f = \sum_i^N m_i
\end{equation}
where $m_i$ is the constituent particle mass.
In the above, $N$ depends on the intrinsic quantum number of the constituent particles:
$N = N_{sd}^{3/2}$.
For example, $N = 1/L$ with the lepton number $L$ for a constituent particle in the formation of a lepton.

The difference number of fermions $N_{sd}$ is the origin of symmetry violation during
DSSB. Fermions with odd parity condense in the vacuum space while fermions with even
parity remain in the matter space; for example, magnetic monopoles with odd parity are
not observed, but electric monopoles with even parity are observed in the matter space. Discrete
symmetries are violated so as to have complex scattering amplitude and the
non-conservation of the right-handed singlet current. This is the main reason for the
change of the fermion mass and gauge boson mass.

\subsection{Fine and Hyperfine Structure}

In order to obtain the fermion mass formula for fine and hyperfine interactions the analogy of QED is nonrelativistically considered
to avoid complexity.
The fine interactions become isospin-isospin interactions as a result of the $SU(2)_L$ gauge theory.
Hyperfine interactions consist of spin-spin and colorspin-colorspin interactions.
The colorspin-colorspin interaction is closely related to the difference between the lepton as a color singlet state and
the quark as a color triplet state.

In QED, the dipole moment has the form expected for a Dirac point-like fermion:
\begin{math}
\vec \mu_i = \frac{e}{2 m_e} \vec \sigma_i
\end{math}
where $e$ is the electric charge of particle, $m_e$ the particle mass, and $\vec \sigma_i$ its Pauli matrix.
The spin-spin interaction due to the magnetic moment leads to the hyperfine splitting of the ground state:
\begin{equation}
\triangle E_{hf} = \frac{2}{3} \vec \mu_i \cdot \vec \mu_j |\psi (0)|^2
= \frac{2 \pi \alpha_e}{3} \frac{\vec \sigma_i \cdot \vec \sigma_j}{m_i m_j} |\psi (0)|^2
\end{equation}
where $\psi (0)$ is the wave function of the two particle system at the origin $(r_{ij} = 0)$
and the spin-spin interaction is proportional to $\vec \sigma_i \cdot \vec \sigma_j = 4 \vec s_i \cdot \vec s_j$.
The above result can be taken over to the isospin-isospin interaction as the fine interaction:
\begin{equation}
\label{fime}
\triangle E_{f}
= \frac{2 \pi g_i^2}{3} \frac{\vec \tau_i \cdot \vec \tau_j}{m_i m_j} |\psi (0)|^2
\end{equation}
where $g_i$ is the isospin coupling constant, $i_f = \vec \tau_i \cdot \vec \tau_j$ is the isospin factor, and $\vec \tau_i = 2 i$ is the Pauli matrix.
The masses $m_i$ and $m_j$ denote constituent fermion masses for each lepton or quark as suggested by the
fermion mass formation through the pairing mechanism.

The above result can be taken over to spin-spin and colorspin-colorspin interactions:
\begin{equation}
\label{hfme}
\triangle E_{hf}
= \frac{2 \pi i_f g_i^2}{3} \frac{\vec \sigma_i \cdot \vec \sigma_j}{m_i m_j} |\psi (0)|^2
+ \frac{2 \pi i_f g_i^2}{3} \frac{\vec \zeta_i \cdot \vec \zeta_j}{m_i m_j} |\psi (0)|^2
\end{equation}
where $g_i$ is the isospin coupling constant, $i_f$ is the
symmetric isospin factor, and $\vec \zeta_i = 2 c$ is the Pauli
matrix. The first term denotes the contribution from the spin-spin
interaction and the second term denotes the colorspin-colorspin
interaction. The second term in (\ref{hfme}) makes the distinction
between leptons and quarks and causes quark mixing like d and s
quarks.

\subsection{Quark and Lepton Mass Generation}

The quark or lepton mass consists of three parts apart from the dual pairing mechanism:
constituent particle mass, the fine structure energy, and the hyperfine structure energy.

Combining (\ref{gafe1}), (\ref{fime}), and (\ref{hfme}), the fermion mass formula is thus
given by
\begin{eqnarray}
\label{fema} m_f & = & \sum_{i} m_i + \frac{2 \pi g_i^2}{3}
\sum_{i >j} \frac{\vec \tau_i \cdot \vec \tau_j}{m_i
m_j} |\psi (0)|^2  \nonumber \\
& + & \frac{2 \pi i_f g_i^2}{3} \sum_{i >j} \frac{\vec \sigma_i
\cdot \vec \sigma_j}{m_i m_j} |\psi (0)|^2 \nonumber \\ & + &
\frac{2 \pi i_f g_i^2}{3} \sum_{i >j} \frac{\vec \zeta_i \cdot
\vec \zeta_j}{m_i m_j} |\psi (0)|^2
\end{eqnarray}
where $|\psi (0)|^2 \simeq (i_f m_f \alpha_i)^3$ and
$m_i$ is the mass of each constituent particle.
The first term of the right-hand side denotes the free constituent particle contribution,
the second term denotes the isospin-isospin contribution, the third term denotes the spin-spin contribution,
and the fourth term denotes the colorspin-colorspin contribution in the mass generation mechanism.
The fourth term seems to be the main reason for the mass difference of the lepton as a color singlet state and the quark as a color triplet state.
The fundamental particles known as leptons and quarks are postulated to be composite particles with the same third component of spin $\pm 1/2$ and
isospin $\pm 1/2$ but a different degeneracy number $N_{sd}$:
$N_{sd}$ is further discussed in the following section.
It suggests that fermion mass generation mechanism is relevant for the (V+A) current anomaly, which
reduces the zero-point energy and triggers the DSSB of gauge and chiral symmetries.

Although the GWS model holds many attractive features, the Higgs
sector and the fermion mass sector are the least satisfactory aspects
of the electroweak theory. The minimal choice of a simple Higgs
doublet is sufficient to generate the masses both of the gauge
bosons and of fermions, but the masses of the fermions are just
parameters of the theory which are not predicted; their empirical
values must be taken as input parameters. The GWS model allows the electron
to be very light but it can not explain why the electron is so
light compared with the intermediate vector boson. However, this
scheme might express a plausible reason for it if the fermion
difference number $N_{sd}$ is very large, about
$10^{12}$ order. QWD hints at mass generation through the $\Theta$
vacuum and the dual pairing mechanism. QWD produces three generations
of fermion families and the gauge bosons for the DSSB of the gauge group.
In this scheme, neutrinos should have masses as hinted at by  \cite{Cows,Fuku}.
QWD as a gauge theory has the favorable feature of only
one input parameter for the weak coupling constant $g_i$; the
dynamics can be regarded as quantum flavordynamics for fermion
generations. Being anomaly free in perturbative renormalization, which
requires equal numbers in quarks and leptons, suggests that the
lepton number excess compared to the baryon number might be used
for the dark matter, which plays an important role in the later
stage of the evolution of the universe. The presence of the
non-perturbative anomaly, the $\Theta$ vacuum term in (\ref{thet}),
does not necessarily spoil renormalization because there is no
Ward-Takahashi identity \cite{Ward} destroyed by the
non-conservation of any local conserved current with even parity.
QWD as the $SU(3)_I$ gauge theory suggests the possibility not for the
separate conservation of baryon number and lepton number
but for the combined of $(B - L)$ number conservation
in the energies above $10^2$ GeV. The lepton number is however
conserved as a result of the $U(1)_Y$ gauge theory below the
weak scale, and the baryon number is conserved as a result of the
$U(1)_Z$ gauge theory below the strong scale \cite{Roh3,Roh31}.

\section{$\Theta$ Constant and Quantum Numbers}
\label{sec6}

The $\Theta$ term in the Lagrangian density triggers DSSB and quantizes space and
time. The parameter $\Theta$ is constrained to hold the flat universe condition. The gauge
invariance and boundary condition of spacetime provide the quantization of the
internal space and external space. The $\Theta$ constant in (\ref{thev}) is
related to the decay of the neutral kaon, and is applied to the mechanism of fermion
mass generation.
In this section, the $\Theta$ constant, matter and vacuum quantum numbers, and $\Theta$ constant and quantum numbers
are addressed.

\subsection{$\Theta$ Constant}

Under the constraint of the extremely flat universe required by quantum gauge theory,
the $\Theta$ constant in (\ref{thet})
\begin{math}
\Theta = 10^{-61} \ \rho_G/\rho_m ,
\end{math}
depends on the gauge boson mass $M_G$ since $\rho_G = M_G^4$:
\begin{math}
\Theta = 10^{-61} \ M_G^4/\rho_c \simeq 10^{-4}
\end{math}
at the weak scale $M_G \simeq 10^{2}$ GeV.
Since the gauge boson mass depends on the Weinberg mixing angle $\theta_W$, $M'_G \rightarrow M_G \sin \theta_W$,
during the DSSB of $SU(3)_I \rightarrow SU(2)_L \times U(1)_Y$ or $SU(2)_L \times U(1)_Y \rightarrow U(1)_e$ gauge theory,
the change of the $\Theta$ constant depends on $\theta_W$:
\begin{math}
\label{thvr}
\Delta \Theta \propto \sin^4 \theta_W = i_f^{e 2}.
\end{math}
Note that the isospin factor $i_f^w = \sin^2 \theta_W \simeq 1/4$ and the weak boson mass $M_W = M_Z \cos \theta_W$.
The relation between the $\Theta$ constant and the difference number $N_{sd}$ is given by
\begin{equation}
\Theta = \pi^2 i_f^4 \alpha_i^4 m_f^4 N_{sd}^2/10^{61} \rho_c
\end{equation}
from equations (\ref{thev}) and (\ref{gafe5}).
$\Theta$ values become
$\Theta_{EW} \approx 10^{-4}$ and $\Theta_{QCD} \approx 10^{-13}$ at different stages.
This is consistent with the observed results, $\Theta < 10^{-9}$ in the electric dipole moment of the neutron
and $\Theta \simeq 10^{-3}$ in the neutral kaon decay.

\subsection{Matter and Vacuum Quantum Numbers}

There is the condensation process in the fermion mass generation mechanism. The process is
the dual pairing mechanism of constituent fermions, which makes boson-like particles of
paired fermions. At the phase transition, $N_{sc}$ becomes zero so that $N_{sd}$
becomes the maximum. Using the relations $M_G = \sqrt{\pi} m_f i_f \alpha_i \sqrt{N_{sd}}$ and $M_G^2 =
M_H^2 - i_f g_i^2 \langle \phi \rangle^2 = i_f g_i^2 [\phi^2 - \langle \phi \rangle^2]$, the zero point energy
$M_{H}^2 = \pi m_f^2 i_f^2 \alpha_i^2 N_{ss}$ and the reduction of the zero-point energy
$\langle \phi \rangle^2 = m_f^2 i_f \alpha_i N_{sc}/4$ are obtained. The difference number of right-left handed singlet
fermions $N_{sd}$ in intrinsic two-space dimensions suggests the introduction of a
degenerate particle number $N_{sp}$ in the intrinsic radial coordinate and an
intrinsic principal number $n_m$; particle quantum numbers are related by
$n_m^4 = N_{sp}^2 = N_{sd}$ and the Dirac quantization condition
$\sqrt{i_f} g_i g_{im} = 2 \pi N_{sp}$ is satisfied.
The $N_{sp}$ is thus the degenerate state number in the
intrinsic radial coordinate that has the same principal number $n_m$. The intrinsic
principal quantum number $n_m$ consists of three quantum numbers, that is, $n_m =
(n_c, n_i, n_s)$ where $n_c$ is the intrinsic principal quantum number for the color
space and, $n_i$ is the intrinsic principal quantum number for the isospin space, $n_s$ is
the intrinsic principal quantum number for the spin space. The intrinsic quantum numbers
$(n_c, n_i, n_s)$ take integer numbers. A fermion therefore possesses a set of
intrinsic quantum numbers $(n_c, n_i, n_s)$ to represent its intrinsic quantum states.

The concept automatically adopts the three types of intrinsic angular momentum
operators, $\hat C$, $\hat I$, and $\hat S$, when the intrinsic potentials for color,
isospin, and spin charges are central so that they depend on the intrinsic radial
distance: for instance, the color potential in the strong interactions is dependent on the
radial distance. The intrinsic spin operator $\hat S$ has a magnitude squared $\langle
S^2 \rangle = s (s + 1)$ and $s = 0, 1/2, 1, 3/2 \cdot \cdot \cdot (n_s-1)$. The third
component of $\hat S$, $\hat S_z$, has half integer or integer quantum number in the
range of $- s \sim s$ with the degeneracy $2s + 1$. The intrinsic isospin operator
$\hat I$ analogously has a magnitude squared $\langle I^2 \rangle = i (i + 1)$ and $i =
0, 1/2, 1, 3/2 \cdot \cdot \cdot (n_i-1)$. The third component of $\hat I$, $\hat
I_z$, has half integer or integer quantum number in the range of $- i \sim i$ with the
degeneracy $2i + 1$. The intrinsic color operator $\hat C$ analogously has a magnitude
square $\langle C^2 \rangle = c (c + 1)$ and $c = 0, 1/2, 1, 3/2 \cdot \cdot \cdot
(n_c-1)$. The third component of $\hat C$, $\hat C_z$, has half integer or integer
quantum number in the range of $- c \sim c$ with the degeneracy $2c + 1$. The
principal number $n_m$ in intrinsic space quantization is very much analogous to the
principal number $n$ in extrinsic space quantization and the intrinsic angular momenta
are analogous to the extrinsic angular momentum so that the total angular momentum has
the form of $\vec J = \vec L + \vec S + \vec I + \vec C$, which is the extension of
the conventional total angular momentum $\vec J = \vec L + \vec S$. The intrinsic
principal number $n_m$ denotes the intrinsic spatial dimension or radial quantization:
$n_c = 3$ represents the strong interactions as an $SU(3)_C$ gauge theory and $n_i = 3$
represents the weak interactions as an $SU(3)_I$ gauge theory. For QWD as the $SU(3)_I$
gauge theory, there are nine weak gauge bosons ($n_i^2 = 3^2 = 9$), which consist of
one singlet gauge boson $G_0$ with $i=0$, three degenerate gauge bosons $G_1 \sim G_3$
with $i=1$, and five degenerate gauge bosons $G_4 \sim G_8$ with $i=2$; for the GWS
model as the $SU(2)_L \times U(1)_Y$ gauge theory, one singlet gauge boson $G_0$ with
$i=0$, three gauge bosons $G_1 \sim G_3$ with $i=1$, and one gauge boson $G_8$ with
$i=2$ are required. One explicit piece of evidence of colorspin and isospin angular momenta is
the strong isospin symmetry in nucleons, which is postulated to be the combination symmetry
of colorspin and weak isospin in this scheme. Other evidence is the nuclear magnetic
dipole moment: the Lande spin g-factors of the proton and neutron are respectively
$g_s^p = 5.59$ and $g_s^n = - 3.83$, which are shifted from $2$ and $0$, because of
contributions from color and isospin degrees of freedom as well as spin degrees of
freedom. The mass ratio of the proton and the constituent quark, $m_p/m_q \sim 2.79$,
thus represents three intrinsic degrees of freedom of color, isospin, and spin. In
fact, the extrinsic angular momentum associated with the intrinsic angular momentum
may be decomposed by $\vec L = \vec L_i + \vec L_c + \vec L_s$ where $\vec L_i$ is the
angular momentum originating from the isospin charge, $\vec L_c$ is the angular
momentum originating from the color charge, and $\vec L_s$ is the angular momentum
originating from the spin charge. This is supported by the fact that the orbital
angular momentum $l_c$ of a nucleon has the different origin from the color charges with
the orbital angular momentum $l_i$ of the electron from the isospin charge, since two
angular momenta have opposite directions from the information of the spin-orbit couplings
in nucleus and atoms. Extrinsic angular momenta have extrinsic parity $(-1)^l =
(-1)^{(l_c + l_i + l_s)}$, intrinsic angular momenta have intrinsic parity $(-1)^{(c +
i + s)}$, and the total parity becomes $(-1)^{(l + c + i + s)}$ for the electric moments;
while the extrinsic angular momenta have extrinsic parity $(-1)^{(l + 1)} = (-1)^{(l_c +
l_i + l_s + 1)}$, intrinsic angular momenta have intrinsic parity $(-1)^{(c + i + s +
1)}$, and the total parity becomes $(-1)^{(l + c + i + s + 1)}$ for magnetic moments.

Fermions increase their masses by
decreasing their intrinsic principal quantum numbers from the higher ones at higher
energies to the lower ones at lower energies. The coupling constant $\alpha_i$ of a
non-Abelian gauge theory is strong for the small $N_{sd}$ and is weak for the large
$N_{sd}$ according to the renormalization group analysis. The vacuum energy is
described by the zero-point energy in the units of $\omega/2$ with the maximum number
$N_{sd} \simeq 10^{61}$ and the vacuum is filled with fermion pairs of up and down
colorspins, isospins, or spins, whose pairs behave like bosons quantized by the unit
of $\omega$: this is analogous to the superconducting state of fermion pairs. The
intrinsic particle number $N_{sp} \simeq 10^{30}$ (or $B \simeq 10^{-12}$, $L \simeq
10^{-9}$) characterizes gravitational interactions for fermions with the mass
$10^{-12}$ GeV, $N_{sp} \simeq 10^{6}$ (or $L_e \simeq 1$) characterizes the weak
interactions for electrons, and $N_{sp} \simeq 1$ (or $B \simeq 1$) characterizes the
strong interactions for nucleons. The fundamental particles known as leptons and quarks
are hence postulated to be composite particles with the color, isospin, and spin quantum
numbers; the quark is a color triplet state, but the lepton is a color singlet state.
Note that if $N_{sp} > 1$ (or $B < 1$), it represents a pointlike fermion and if
$N_{sp} < 1$ (or $B > 1$), it represents a composite fermion. There is accordingly the
possibility of internal structure for the lepton or quark when they are considered as
composite particles. Since the $\Theta$ constant is thus parameterized by $\Theta =
10^{-61} \ \rho_G/\rho_m$ with the vacuum energy density $\rho_G = M_G^4$ and the
matter energy density $\rho_m \simeq \rho_c \simeq 10^{-47} \ \textup{GeV}^4$, the
relation between the $\Theta$ constant and the difference number $N_{sd}$ is given by
$\Theta = \pi^2 m_f^4 i_f^4 \alpha_i^4 N_{sd}^2/10^{61} \rho_c$.

The weak vacuum represented by massive gauge bosons is quantized by the maximum wavevector mode
$N_R \approx 10^{30}$, the total gauge boson number $N_G = 4 \pi N_R^3/3 \approx 10^{91}$,
and the gauge boson number density $n_G = \Lambda_{EW}^3 \approx 10^{47} \ \textup{cm}^{-3}$.
Baryon matter represented by massive baryons is spatially quantized by the maximum wavevector mode
$N_F \approx 10^{26}$ and the total baryon number $B = N_B = 4 \pi N_F^3/3 \approx 10^{78}$  \cite{Roh31,Roh11}.
The baryon matter quantization described above is consistent with the
nuclear matter density $n_n \approx n_B \approx 1.95 \times 10^{38} \ \textup{cm}^{-3}$ and
Avogadro's number $N_A = 6.02 \times 10^{23} \ \textup{mol}^{-1} \approx 10^{19} \ \textup{cm}^{-3}$ in the matter.
Electrons with the mass $0.5$ MeV might be similarly quantized by $N_F \approx 10^{27}$ and the total number $10^{81}$
if the electron number is conserved under the assumption of $\Omega_e = \rho_e/\rho_c \approx 1$:
since the conservation of the baryon number minus the lepton number ($B - L$) as well as the baryon number ($B$)
and the lepton number ($L$) is a good quantum number at low energies,
the total electron number $10^{81}$, different from the total baryon number $10^{78}$, suggests lepton matter as dark matter.
The maximum wavevector mode $N_F$ is close to $10^{30}$
if the quantization unit of fermions $10^{-12}$ GeV (the baryon number $B \simeq 10^{-12}$) is used rather than the unit of baryons $0.94$ GeV
(the baryon number $B = 1$) under the assumption of the fermion number conservation $N_f \simeq 10^{91}$.

\subsection{$\Theta$ Constant and Quantum Numbers}

The invariance of the gauge transformation provides us with
$\psi [\hat O_\nu] = e^{i \nu \Theta} \psi [\hat O]$ for the fermion wave function $\psi$ with the transformation of an operator $\hat O$ by
the class $\nu$ gauge transformation, $\hat O_\nu$:
the vacuum state characterized by the constant $\Theta$ is called the $\Theta$ vacuum \cite{Hoof2}.
The true vacuum is the superposition of all the $|\nu \rangle$ vacua with the phase $e^{i \nu \Theta}$:
$|\Theta \rangle = \sum_\nu e^{i \nu \Theta} |\nu \rangle$.
The topological winding number $\nu$ or the topological charge $q_s$ is defined by
\begin{equation}
\label{tonu}
\nu = \nu_+ - \nu_- = \int \frac{i_f g_i^2}{16 \pi^2} Tr F^{\mu \nu} \tilde F_{\mu \nu} d^4 x
\end{equation}
where the subscripts $+$ and $-$ denote moving particles with chiralities $+$ and $-$ respectively in the presence of the gauge fields \cite{Atiy}.
The matter energy density generated by the surface effect is postulated to be
\begin{equation}
\label{toma}
\rho_{m} \simeq \rho_c \simeq \frac{i_f g_i^2}{16 \pi^2} Tr F^{\mu
\nu} \tilde F_{\mu \nu}
\end{equation}
which implies that the fermion mass is generated by the difference of fermion numbers
moving to left- and right-handed directions. In this respect, the difference number
$N_{sd}$, the left-handed fermion number $N_{ss}$, and the condensed right-handed fermion
number $N_{sc}$ in intrinsic two-space dimensions respectively correspond to $\nu$,
$\nu_+$, and $\nu_-$ in three-space and one time dimensions.  In the presence of the $\Theta$
term, the singlet $(V+A)$ current is not conserved due to an Adler-Bell-Jackiw anomaly
\cite{Adle}:
\begin{math}
\partial_\mu J_\mu^{1 + \gamma^5} = \frac{N_f i_f g_i^2}{16 \pi^2} Tr F^{\mu \nu} \tilde F_{\mu \nu}
\end{math}
with the flavor number of fermions $N_f$, and this reflects
degenerate multiple vacua. This illustrates mass generation by
the surface effect due to the field configurations with parallel
isospin electric and magnetic fields.
If $\nu = \rho_m/\rho_G$ is introduced from (\ref{tonu}) and (\ref{toma}),
a condition $\Theta \nu = 10^{-61}$ is satisfied.
The $\Theta$ value parameterized by
$\Theta = 10^{-61} \rho_G/\rho_m$ is consistent with the observed
result, $\Theta \simeq 10^{-3}$, in the neutral kaon decay \cite{Chri}.

The topological winding number $\nu$ is related to the intrinsic quantum number $n_m$ by $\nu = n_m^{-8}$.
The intrinsic principal number $n_m$ is also connected with $N_{sp}$ and $N_{sd}$:
$n_m^2 = N_{sp}$, $N_{sp}^2 = N_{sd}$, and $N_{sp}^{4} = 1/\nu$.
The relation between the intrinsic radius and the intrinsic quantum number might be written by
\begin{math}
r_i = r_{0i} / n_m^2
\end{math},
with the radius $r_{0i} = 1/2 m_f \alpha_z = N_{sp}/M_G$. The intrinsic
quantum numbers are exactly analogous to the extrinsic quantum
numbers. The extrinsic principal number $n$ for the nucleon is
related to the nuclear mass number $A$ or the baryon quantum
number $B$: $n^2 = A^{1/3}$, $n^4 = A^{2/3}$, $n^6 = B = A$. The
relation between the nuclear radius and the extrinsic quantum
number is outlined by
\begin{equation}
r = r_0 A^{1/3} = r_0 n^2
\end{equation}
with the radius $r_0 = 1.2$ fm and the nuclear principal number $n$. This is analogous
to the atomic radius $r_e = r_0 n_e^2$ with the atomic radius $r_0$ and the electric
principal number $n_e$: the atomic radius $r_0 = 1/2 m_e \alpha_y$ is almost the same
as the Bohr radius $a_B = 1/m_e \alpha_e = 0.5 \times 10^{-8}$ cm. These concepts
are related to the constant nuclear density $n_B = 3/4 \pi r_0^3 = 1.95 \times 10^{38}
\ \textup{cm}^{-3}$ or Avogadro's number $N_A = 6.02 \times 10^{23} \
\textup{mol}^{-1}$ and to the constant electron density $n_e = 3/4 \pi r_e^3 = 6.02
\times 10^{23} Z \rho_m/A$ with the nuclear mass density $\rho_m$ in the unit of $\textup{g/cm}^3$,
where the possible relation is
\begin{equation}
r_e = r_0 L^{1/3} = r_0 n_e^2
\end{equation}
with the lepton number $L$.

$\Theta$ values according to (\ref{thev}) become $\Theta_{Pl}
\approx 10^{61}$, $\Theta_{EW} \approx 10^{-4}$, $\Theta_{QCD}
\approx 10^{-13}$, and $\Theta_{0} \approx 10^{-61}$ at different
stages. The scope of $\Theta = 10^{61} \sim 10^{-61}$ corresponds
to the scope of $\nu = 10^{-122} \sim 10^{0}$ to satisfy the flat
universe condition $\nu \Theta = 10^{-61}$: the maximum
quantization number $N_{sp} \simeq N_R \simeq 10^{30}$ and $N_G
\simeq 4 \pi N_R^3/3 \simeq 10^{91}$. The maximum wavevector mode
$N_R = (\rho_G/\Theta \rho_m)^{1/2} = 10^{30}$ of the weak vacuum
is obtained. These describe possible dualities between intrinsic
quantum numbers and extrinsic quantum numbers: $n_m$ and $n$,
$N_{sp}^3$ and $A$, and $1/\nu$ and $A^{4/3}$ for baryons.
$\Theta$ term as the surface term modifies the original GWS model
\cite{Glas} for the weak interactions, which has a problem in the
fermion mass violating gauge invariance, and this suggests mass
generation as the non-perturbative breaking of gauge and chiral
invariance through DSSB.

\section{Comparison among Quantum Weakdynamics, Glashow-Weinberg-Salam model,
and Grand Unified Theory} \label{sec7}

This section is devoted to a summary, and to show QWD extending beyond the SM or toward a new
GUT as an $SU(3)_C \times SU(3)_I$ gauge theory. Table \ref{comp} summarizes a
comparison among QWD, GWS model, and GUT. GUT in the table represents $SU(5)$ gauge
theory or $SO(10)$ gauge theory \cite{Geor}.

There are many unexplained phenomena in the SM.
For instances, there are no clues in the SM for many
free parameters, three family generations for the leptons and quarks, matter mass
generation, the Higgs problem or vacuum problem, dynamical spontaneous symmetry breaking
(DSSB) beyond spontaneous symmetry breaking, the neutrino mass problem, etc. In order to
resolve these problems, grand unified theories (GUTs) were proposed \cite{Geor}.
Nevertheless, the grand unification of strong and electroweak interactions is not
complete and GUTs also have model dependent problems: hierarchy problem, proton decay,
and the Weinberg angle are problems in $SU(5)$ gauge theory \cite{Geor}.

Already confirmed predictions of QWD generating the GWS model are
discrete symmetry breaking, V - A current conservation, massive
gauge bosons, the Fermi weak coupling constant $G_F$, the Weinberg
angle $\sin^2 \theta_W = 1/4$, the Cabbibo angle $\sin \theta_C =
1/4$, and the three generations of leptons and quarks as discussed in
Section \ref{sec3}.

New predictions from QWD beyond the GWS model, which are to be tested by
experiments, are as follows: the existence of isospin $3/2$ leptons or quarks in
addition to isospin $1/2$ fermions, the eight-fold way of isospin $1/2$ fermions as
composite particles, CP violation with $\Theta \simeq 10^{-4}$ order, the existence of
massive gauge bosons $G_4 \sim G_7$ with isospin $2$ in addition to gauge bosons $G_1 \sim G_3$ with
isospin $1$, the (B-L) current conservation above the weak scale, the new unification
scale of the strong and weak forces at the order of $10^{3}$ GeV, fermion mass generation,
neutrino mass and oscillation confirmed by recent experiments,
quark flavor mixing due to color charge, the isospin coupling constant
hierarchy ($\alpha_i, \ \alpha_z = \alpha_i/3, \ \alpha_w = \alpha_w/4, \ \alpha_y =
\alpha_i/12$, and $\alpha_e = \alpha_i/16$), the lepton asymmetry in the universe,
neutron-antineutron oscillations, the analogy between QWD and QCD as $SU(3)$ gauge theories, etc.

\section{Conclusions}
\label{sec8}

This study proposes that a certain group $H$ leads to an $SU(3)_I
\times SU(3)_C$ group for weak and strong interactions at a grand
unification scale through dynamical spontaneous symmetry breaking
(DSSB); the group chain is $H \supset SU(3)_I \times SU(3)_C$. The
grand unification of the $SU(3)_I \times SU(3)_C$ group beyond the
standard model with the group $SU(3)_C \times SU(2)_L \times
U(1)_Y$ provides the coupling constant $\alpha_i = \alpha_s
\simeq 0.12$ at a new grand unification scale around $10^{3}$
GeV, which might be the resolution to the hierarchy problem of
the grand unification scale $10^{15}$ GeV. DSSB consists of two
simultaneous mechanisms; the first mechanism is the explicit
symmetry breaking of gauge symmetry, which is represented by the
isospin (isotope) factor $i_f$ and the weak coupling constant
$g_i$, and the second mechanism is the spontaneous symmetry
breaking of gauge fields, which is represented by the
condensation of the pseudo-scalar fields postulated as spatially
longitudinal components of the gauge fields. At the energy $T \simeq
10^2$ GeV, the gauge boson number density $n_G = M_G^3 \approx
10^6 \ \textup{GeV}^3 \approx 10^{47} \ \textup{cm}^{-3}$, the
vacuum energy density $V_0 = M_G^4 \approx 10^8 \ \textup{GeV}^4
\approx 10^{25} \ \textup{g \ cm}^{-3}$, and the total gauge
boson number $N_G \approx 10^{91}$ are predicted.

Quantum weakdynamics (QWD) as an $SU(3)_I$ gauge theory predicts
the free parameters in the GWS model being an $SU(2)_L \times
U(1)_Y$ gauge theory. QWD provides the Weinberg angle, the
Cabbibo angle, quark and lepton families, a modification of the
Higgs mechanism, fermion mass generation, etc. QWD is dynamically
spontaneous symmetry broken through the condensation of pseudo-
scalar bosons. QWD generates electroweak theory as an $SU(2)_L
\times U(1)_Y$ gauge theory at the electroweak scale and then
electroweak theory produces QED as a $U(1)_e$ gauge theory
through DSSB. The essential point is that the DSSB mechanism is
adopted to all the interactions characterized by gauge
invariance, the physical vacuum problem, and discrete symmetry
breaking. This work suggests that the electroweak interactions
originate from the $SU(3)_I$ gauge theory for the weak force,
provided that scalar bosons play the roles of Higgs particles
producing DSSB. The effective coupling constant chain due to
massive gauge bosons is $G_H \supset G_F \times G_R$ for grand
unification, weak, and strong interactions respectively. Quark
and lepton families seems to be successfully generated in terms
of the mixing of the $SU(2)$ three subgroups in two octets. The
DSSB of local gauge symmetry and global chiral symmetry triggers
the (V+A) current anomaly. The (V - A) current conservation and
the (V+A) current non-conservation are thus explained and the
absence of right-handed neutrinos is resulte of the (V+A)
current non-conservation. The electroweak coupling constants derived
in terms of the Weinberg angle $\sin^2 \theta_W = 1/4$ are
$\alpha_z = i_f^z \alpha_i = \alpha_i/3 \simeq 0.04$, $\alpha_w =
i_f^w \alpha_i = \alpha_i/4 \simeq 0.03$, $\alpha_y = i_f^y
\alpha_i = \alpha_i/12 \simeq 0.01$, and $\alpha_e = i_f^e
\alpha_i = \alpha_i/16 \simeq 1/133$ for symmetric isospin
interactions at the weak scale and $-2 \alpha_i/3$, $-
\alpha_i/2$, $- \alpha_i/6$, and $- \alpha_i/8$ for asymmetric
isospin interactions. The isospin factors are $i^s_f = (i_f^z,
i_f^w, i_f^y, i_f^e) = (1/3, 1/4, 1/12, 1/16)$ for symmetric
interactions and $i^a_f = (-2/3, -1/2, -1/6, -1/8)$ for
asymmetric interactions: $i_f^w = \sin^2 \theta_W$ and $i_f^e =
\sin^4 \theta_W$. The symmetric charge factors reflect an intrinsic
even parity with repulsive force while the asymmetric charge
factors reflect intrinsic odd parity with attractive force; this
suggests electromagnetic duality. The Cabbibo angle $\sin
\theta_C \simeq 1/4$ predicts that two more massive gauge bosons
with $M_G = 2 M_W$ in addition to four intermediate vector bosons
exist at the weak scale for symmetric isospin interactions and
quark flavor mixing is due to color degrees of freedom for
quarks.  QWD for weak interactions is also proposed as the
analogous dynamics of QCD for strong interactions. Common
characteristics of gauge theories, such as gauge symmetry, the true
vacuum problem, and discrete symmetry breaking for both weak and
strong interactions, are understood in terms of the concepts of
the analogy property and DSSB mechanism.

The mechanism of fermion mass generation is suggested in terms of the DSSB of gauge
symmetry and chiral symmetry known as the dual pairing mechanism of the
superconducting state: $M_G = \sqrt{\pi} m_f i_f \alpha_i \sqrt{N_{sd}}$ with
the difference number of right-left-handed fermions $N_{sd}$ in intrinsic two-space dimensions.
The $\Theta$ constant is parameterized by
$\Theta = 10^{-61} \ \rho_G/\rho_m$ with the vacuum energy density $\rho_G = M_G^4$
and the matter energy density $\rho_m$.   $N_{sd}$ suggests the introduction of a degenerate particle
number $N_{sp}$ in the intrinsic radial coordinate and an intrinsic principal number
$n_m$; particle numbers are linked to the relation $n_m^4 = N_{sp}^2 = N_{sd}$ and the
Dirac quantization condition $\sqrt{i_f} g_i g_{im} = 2 \pi N_{sp}$ is satisfied. The
intrinsic principal quantum number $n_m$ consists of three quantum numbers, that is,
$n_m = (n_c, n_i, n_s)$ where $n_c$ is the intrinsic principal quantum number for the
color space, $n_i$ is the intrinsic principal quantum number for the isospin space and
$n_s$ is the intrinsic principal quantum number for the spin space.
The concept automatically adopts the three types of intrinsic angular momentum
operators, $\hat C$, $\hat I$, and $\hat S$, when intrinsic potentials for color,
isospin, and spin charges are central so that they depend on the intrinsic radial
distance. The principal number $n_m$ in intrinsic space quantization is very
much analogous to the principal number $n$ in extrinsic space quantization and the
intrinsic angular momenta are analogous to the extrinsic angular momentum so that the
total angular momentum has the form of $\vec J = \vec L + \vec S + \vec I + \vec C$,
which is the extension of the conventional total angular momentum $\vec J = \vec L +
\vec S$.  The intrinsic particle number $N_{sp} \simeq 10^{6}$ (or $L_e \simeq 1$)
characterizes weak interactions for electrons, and $N_{sp} \simeq 1$ (or $B \simeq 1$) characterizes
strong interactions for nucleons. Fundamental particles known as leptons and quarks
are hence postulated as composite particles with the color, isospin, and spin quantum
numbers; the quark is a color triplet state but the lepton is a color singlet state.
If $N_{sp} > 1$ (or $B < 1$), it represents a point-like fermion and if
$N_{sp} < 1$ (or $B > 1$), it represents a composite fermion.
The $\Theta$ value defined by $\Theta = 10^{-61} \rho_G/\rho_m$ is consistent with the observed result,
$\Theta \simeq 10^{-3}$ in the neutral kaon decay.
The fact that a gauge theory allows one to generate the masses of fermions and gauge bosons without
spoiling the renormalizability is very important;
the renormalizability of a gauge theory with spontaneous symmetry breaking was demonstrated
by 't Hooft \cite{Hoof}.

In this scheme, vacuum and matter energies are spatially quantized as well as photon energy.
The weak vacuum represented by massive gauge bosons is quantized by the maximum wavevector mode
$N_R \approx 10^{30}$, the total gauge boson number $N_G = 4 \pi N_R^3/3 \approx 10^{91}$,
and the gauge boson number density $n_G = \Lambda_{EW}^3 \approx 10^{47} \ \textup{cm}^{-3}$.
The baryon matter represented by massive baryons is quantized by the maximum wavevector mode (Fermi mode)
$N_F \approx 10^{26}$ and the total baryon number $B = N_B = 4 \pi N_F^3/3 \approx 10^{78}$.
Electrons might be similarly quantized by the maximum wavevector mode $N_F \approx 10^{27}$
and the total particle number $L_e = N_e \approx 10^{81}$ for electrons with the mass $0.5$ MeV.
The maximum wavevector mode $N_F$ is close to $10^{30}$
if the quantization unit of fermions $10^{-12}$ GeV is used rather than the unit of baryons, $0.94$ GeV
under the assumption of the fermion number conservation.
The baryon matter quantization described above is consistent with the
nuclear matter density $n_n \approx n_B \approx 1.95 \times 10^{38} \ \textup{cm}^{-3}$ and
Avogadro's number $N_A = 6.02 \times 10^{23} \ \textup{mol}^{-1} \approx 10^{19} \ \textup{cm}^{-3}$ in the matter.
Massless photons are quantized by the maximum wavevector mode
$N_\gamma \approx 10^{29}$ and the total photon number $N_{t \gamma} = 4 \pi N_\gamma^3/3 \approx 10^{88}$.
These total particle numbers $N_G \approx 10^{91}, N_B \approx 10^{78}$, and $N_{t \gamma} \approx 10^{88}$ are conserved good quantum numbers.
The quantization unit of vacuum energy due to a gauge boson in the weak interactions is $\Lambda_{EW}/N_R \simeq 10^{-28}$ GeV.

The notable accomplishments of this work are summarized in the
following. The grand unification of the $SU(3)_I \times SU(3)_C$
gauge theory beyond the standard model of the $SU(3)_C \times
SU(2)_L \times U(1)_Y$ gauge theory is suggested in the scale
around $10^{3}$ GeV rather than in the conventional unification
scale $10^{15}$ GeV. This seems to be the viable solution of the
hierarchy problem. The unification of the $SU(2)_L \times U(1)_Y$
electroweak theory, the GWS model, is developed in terms of the
$SU(3)_I$ gauge theory, QWD, if scalar bosons postulated to be
spatially longitudinal components of the gauge bosons play roles of
Higgs particles during DSSB. The DSSB of local gauge symmetry and
global chiral symmetry gives rise to the (V+A) current anomaly.
The quark and lepton families are successfully generated from two
octets of triplet isospin combinations, and the predicted
Weinberg angle and Cabbibo angle are in good agreement with
experimental values. Photons are regarded as massless gauge
bosons indicating DSSB and the fine structure constant in the
electromagnetic interactions is related to the fine structure
constant in weak interactions, $\alpha_e = \alpha_i /16$. Fermion
mass generation is suggested in terms of the DSSB of gauge
symmetry and chiral symmetry due to the (V+A) current anomaly.
The baryon minus lepton number (B -L) conservation is a
consequence of the $SU(3)_I$ gauge theory, the lepton number
conservation is a consequence of the $U(1)_Y$ gauge theory, and
the electron number conservation is a consequence of the
$U(1)_e$ gauge theory. The development of this theory would thus
shed light on understanding the fundamental forces in nature and its
consequences play significant roles in various fields since
point-like particles, like quarks and leptons are more or less
governed by the electroweak force explained by QWD.

\newpage
\widetext

\begin{table}
\caption{\label{isqu} Weak isospin and hypercharge quantum numbers
of leptons and quarks}
\end{table}
\centerline{
\begin{tabular}{|c|c|c|c|c|} \hline
Leptons  & $I$ & $I_3$ & $Y$ & $Q$ \\ \hline \hline $\nu_e$ &
$1/2$ & $1/2$ & $-1$ & $0$ \\ \hline $e_L$ & $1/2$ & $-1/2$ &
$-1$ & $-1$ \\ \hline $e_R$ & $0$ & $0$ & $-2$ & $-1$ \\ \hline
\end{tabular}
}
\vspace{0.15in}
\centerline{
\begin{tabular}{|c|c|c|c|c|} \hline
Quarks  & $I$ & $I_3$ & $Y$ & $Q$ \\ \hline \hline $u_L$ & $1/2$
& $1/2$ & $1/3$ & $2/3$ \\ \hline $d_L$ & $1/2$ & $-1/2$ & $1/3$
& $-1/3$ \\ \hline $u_R$ & $0$ & $0$ & $4/3$ & $2/3$ \\ \hline
$d_R$ & $0$ & $0$ & $-2/3$ & $-1/3$ \\ \hline
\end{tabular}
} 

\vspace{0.15in}

\begin{table}
\caption{\label{coga} Relations between Conservation Laws and
Gauge Theories}
\end{table}
\centerline{
\begin{tabular}{|c|c|c|} \hline
Force & Conservation Law & Gauge Theory \\ \hline \hline
Electromagnetic & Electron & $U(1)_e$ \\ \hline Weak & Lepton (or
B-L) & $U(1)_Y$ \\ \hline Weak & V-A & $SU(2)_L \times U(1)_Y$ \\
\hline Weak & isotope (isospin) & $SU(3)_I$ \\ \hline
\end{tabular}
}

\vspace{0.15in}

\begin{table*}
\caption{\label{comp} Comparison among Quantum Weakdynamics,
Glashow-Weinberg-Salam model, and Grand Unified Theory}
\end{table*}
\centerline{
\begin{tabular}{|c|c|c|c|} \hline
Classification & QWD & GWS & GUT \\ \hline \hline Grand
Unification energy & $10^{3}$ GeV & & $10^{15}$ GeV \\ \hline
Symmetry breaking & DSSB & SSB & SSB \\ \hline Discrete symmetries
(P, C, T, CP) & breaking &  breaking & breaking \\ \hline $\Theta$
vacuum & yes & no & no \\ \hline Higgs bosons & no & yes & yes \\
\hline Lepton number conservation & yes & yes & no \\ \hline
Electron number conservation & yes ($N_e \simeq 10^{81}$) & yes &
yes \\ \hline Baryon number conservation
& no & no & no \\ \hline Proton decay & unknown & no & yes \\
\hline Electron decay & unknown & no & unknown \\ \hline Weinberg
angle & $\sin^2 \theta_W = 1/4$ & free parameter & $\sin^2
\theta_W = 3/8$  \\ \hline Cabbibo angle & $\sin \theta_C = 1/4$ &
free parameter & free parameter
\\ \hline Fermion mass generation & yes & unsatisfactory & unsatisfactory \\
\hline Coupling constant hierarchy & yes & no & no \\ \hline
Neutrino mass & yes & no & unknown \\ \hline Number of gauge
bosons & 9 & 4  & 24 \\ \hline Free parameters & coupling
constant & many  & several \\ \hline
\end{tabular}
}


\begin{thebibliography}{}
\bibitem{Glas} S. L. Glashow, Nucl. Phys. {\bf 22}, 579 (1961);
S. Weinberg, Phys. Rev. Lett. {\bf 19}, 1264 (1967);
A. Salam, Elementary Particle Theory, N. Svaratholm Ed., Almquist and Wiksells (1968).
\bibitem{Frit} H. Fritzsch and M. Gell-Mann, in Proceedings of the Sixteenth International Conference on High Energy Physics, Vol. {\bf 2}, Chicago (1972);
H. Fritzsch, M. Gell-Mann and Leutwyler, Phys. Lett. {\bf  B 47}, 365 (1973); S.
Weinberg, Phys. Rev. Lett. {\bf 31}, 494 (1973).
\bibitem{Geor} H. Georgi and S. L. Glashow, Phys. Rev. Lett. {\bf 32}, 438 (1974);
H. Georgi, in Particles and Fields, C. E. Carlson Ed., A.I.P. (1975);
H. Fritzsch and P. Minkowski, Ann. Phys. {\bf 93}, 193 (1975).
\bibitem{Hinc} I. Hinchliffe, Phys. Rev. {\bf D 50}, 1297 (1994).
\bibitem{Bard} M. Bardon et al., Phys. Rev. Lett. {\bf 14}, 449 (1965);
A. Sirlin, Phys. Rev. {\bf D 29}, 89 (1984).
\bibitem{Roh3} H. Roh, ``QCD Confinement Mechanism and $\Theta$ Vacuum:
Dynamical Spontaneous Symmetry Breaking,'' hep-th/0012261.
\bibitem{Roh31} H. Roh, ``Quantum Nucleardynamics as an $SU(2)_N \times U(1)_Z$ Gauge Theory,'' nucl-th/0101001.
\bibitem{Higg} P. W. Higgs, Phys. Rev. Lett. {\bf 13}, 508 (1964).
\bibitem{Namb} Y. Nambu, Phys. Rev. Lett. {\bf 4}, 380 (1960);
J. Goldstone, Nuovo Cimento {\bf 19}, 154 (1961).
\bibitem{Suss} L. Susskind, Phys. Rev. {\bf D 20}, 2619 (1979);
S. Weinberg, Phys. Rev. {\bf D 19}, 1277 (1979).
\bibitem{Gell} M. Gell-Mann, Phys. Lett. {\bf 8}, 214 (1964);
K. Nishijima and T. Nakano, Prog. Theo. Phys. {\bf 10}, 581 (1953).
\bibitem{Adle} S. L. Adler, Phys. Rev. {\bf 177}, 2426 (1969);
J. S. Bell and R. Jackiw, Nuovo Cimento {\bf A 60}, 47 (1969).
\bibitem{Hoof2} G. 't Hooft, Phys. Rev. Lett. {\bf 37}, 8 (1976); Phys. Rev. {\bf D 14}, 3432 (1976);
R. Jackiw and C. Rebbi, Phys. Rev. Lett. {\bf 37}, 172 (1976);
C. G. Callan, R. Dashen, and D. Gross, Phys. Lett. {\bf 63 B}, 334 (1976).
\bibitem{Lee} T. D. Lee and C. N. Yang, Phys. Rev. {\bf 104}, 254 (1956).
\bibitem{Koba} M. Kobayashi and K. Maskawa, Prog. Theor. Phys. {\bf 49}, 652 (1973).
\bibitem{Chri} J. H. Christensen, J. W. Cronin, V. L. Fitch, and R. Turlay, Phys. Rev. Lett.
{\bf 13}, 138 (1964).
\bibitem{Murt} S. Murthy et al., Phys. Rev. Lett. {\bf 63}, 965 (1989).
\bibitem{Roh1} H. Roh, ``Toward Quantum Gravity I: Newton Gravitation Constant, Cosmological Constant,
and Classical Tests,'' gr-qc/0101001.
\bibitem{Roh11} H. Roh, ``Toward Quantum Gravity II: Quantum Tests,'' gr-qc/0101002.
\bibitem{Kim} J. E. Kim, P. Langacker, M. Levine, and H. H. Williams, Rev. Mod. Phys. {\bf 53}, 211 (1981).
\bibitem{Hung} P. Q. Hung and J. J. Sakurai, Ann. Rev. Nucl. Particle. Sci. {\bf 31}, 375 (1981).
\bibitem{Pati0} J. C. Pati, A. Salam, and J. Strathdee, Phys. Lett. {\bf B 59}, 265 (1975).
\bibitem{Pati} J. C. Pati and A. Salam, Phys. Rev. {\bf D 10}, 275 (1974);
R. N. Mohapatra and J. C. Pati, Phys. Rev. {\bf D 1}, 566, 2558 (1975);
G. Senjanovic and R. N. Mohapatra, Phys. Rev. {\bf D 12}, 1502 (1975).
\bibitem{Cabb} N. Cabbibo, Phys. Rev. Lett. {\bf 10}, 531 (1963).
\bibitem{Glas1} S. L. Glashow, J. Iliopoulos, and L. Maiani, Phys. Rev. {\bf D 2}, 1285 (1970).
\bibitem{Wein7} S. Weinberg, Tnans. N. Y. Acad. Sci. II, 38 (1977).
\bibitem{Frit2} H. Fritzsch, Phys. Lett. {\bf B 73}, 317 (1978); ibid. {\bf 73}, 3171 (1978).
\bibitem{Dira} P. A. M. Dirac, Proc. of the Royal Soc. {\bf A 133}, 60 (1931).
\bibitem{Poma} A. Pomarol, Phys. Rev. Lett. {\bf 85}, 4004 (2000).
\bibitem{Gros} D. Gross and F. Wilczek, Phys. Rev. Lett. {\bf 30}, 1343 (1973);
Phys. Rev. {\bf D 8}, 3633 (1973); H. D. Politzer, Phys. Rev. Lett. {\bf 30}, 1346
(1973); Phys. Rep. {\bf 14c}, 129 (1974).
\bibitem{Aitc} See, e.g., I. J. R. Aitchison and A. J. G. Hey, {\em Gauge Theories in Particle Physics},
2nd Ed., Adam Hilger, Bristol and Philadelphia, pp 393 - 421
(1989).
\bibitem{Mand} S. Mandelstam, Phys. Rep. {\bf 23c}, 245 (1976);
G. 't Hooft, High Energy Physics, Proc. European Phys. Soc. Int. Conf., A. Zichichi Ed., 1225 (1976).
\bibitem{Seib} N. Seiberg and E. Witten, Nucl. Phys. {\bf B 426}, 19 (1994); ibid. {\bf 431}, 484 (1994);
N. Seiberg, Phys. Rev. {\bf D 49}, 6857 (1994);
K. Intriligator, R. G. Leigh, and N. Seiberg, Phys. Rev. {\bf D 50}, 1092 (1994).
\bibitem{Appe} T. Appelquist and J. Carazzone, Phys. Rev. {\bf D 11}, 2856 (1975).
\bibitem{Cows} R. Cowsik and J. McClelland, Phys. Rev. Lett. {\bf 29}, 669 (1972).
\bibitem{Fuku} Y. Fukuda et al., Phys. Rev. Lett. {\bf 81}, 1562 (1998).
\bibitem{Ward} J. C. Ward, Phys. Rev. {\bf 78}, 1824 (1950);
Y. Takahashi, Nuovo Cimento {\bf 6}, 370 (1957).
\bibitem{Atiy} M. Atiyah and I. Singer, Ann. Math. {\bf 87}, 484 (1968).
\bibitem{Hoof} G. 't Hooft, Nucl. Phys.  {\bf B 33}, 173 (1971); ibid. {\bf B 35}, 167 (1971).
\end{thebibliography}
\end{document}